\documentclass{emulateapj}
\usepackage{amssymb,latexsym,verbatim,graphicx}
\input psfig.sty

\slugcomment{ApJ, in press}

\shorttitle{Generalized Secular Theory}
\shortauthors{Veras \& Armitage}

\begin{document}

\title{Extrasolar planetary dynamics with a generalized planar 
Laplace-Lagrange secular theory}

\author{Dimitri Veras\altaffilmark{1,2} and Philip J. Armitage\altaffilmark{1,2}}
\altaffiltext{1}{JILA, Campus Box 440, University of Colorado, Boulder CO 80309-0440; 
dimitri.veras@colorado.edu; pja@jilau1.colorado.edu}
\altaffiltext{2}{Department of Astrophysical and Planetary Sciences, 
University of Colorado, Boulder CO 80309-0391}

\clearpage
\pagebreak
\newpage

\begin{abstract}
The dynamical evolution of nearly half of the known extrasolar planets 
in multiple-planet systems may be dominated by secular perturbations.  The 
commonly high eccentricities of the planetary orbits calls 
into question the utility of the traditional Laplace-Lagrange (LL)
secular theory in analyses of the motion.  We analytically 
generalize this theory to 
fourth-order in the eccentricities, compare the result with the 
second-order theory and octupole-level theory, and apply these theories to  
the likely secularly-dominated HD 12661, HD 168443, HD 38529 and
$\upsilon$ And multi-planet systems.
The fourth-order scheme yields a multiply-branched criterion 
for maintaining apsidal libration, and implies that the apsidal 
rate of a small body is a function of its initial eccentricity, 
dependencies which are absent from the traditional theory.
Numerical results indicate that the primary difference 
the second and fourth-order theories reveal is an alteration 
in secular periodicities, and
to a smaller extent amplitudes of the planetary eccentricity variation.
Comparison with numerical integrations indicates that the
improvement afforded by the fourth-order theory over the second-order
theory sometimes dwarfs the improvement needed to reproduce the actual
dynamical evolution.  We conclude that LL secular 
theory, to any order, generally represents a poor barometer for 
predicting secular dynamics in extrasolar planetary systems, but does 
embody a useful tool for extracting an accurate long-term dynamical 
description of systems with small bodies and/or near-circular orbits.
\end{abstract}


\section{Introduction}

Dynamical systems of three or more bodies typically include
physical phenomena known as {\it mean motion resonances} and
{\it secular perturbations}.  The former occur when pairs of
bodies have orbital periods whose ratio can be approximately
expressed as a ratio of two small integers.  Otherwise, secular
perturbations dominate the system's long-term evolution.  Each phenomenon
can be described by particular terms in a gravitational potential
which has become commonly known as a ``disturbing function'' 
\citep{ellis00}, a dynamically rich segment of the Hamiltonian of 
the system \citep{morby02}.  Laplace-Lagrange (LL) secular theory, 
reviewed for example
in \cite{murray99}, represents an elegant 
formulation in which the time evolution of objects' 
orbital parameters can be described analytically by 
utilizing a select few terms in the disturbing function up to
second-order in eccentricities and inclinations.

However, the theory has come under recent scrutiny for its failure
to reproduce the phase space structure of dynamical systems of 
interest, such as extrasolar planetary systems.
\cite{libert05,libert06}
showcase the limitations of the theory by using an alternative, 
12th-order expansion in eccentricity, and \cite{beauge06}
successfully reproduce motions of irregular satellites with 
eccentricities as high as $0.7$ by using a third-order Hori
averaging method.  \cite{barnes06}  find significant 
discrepancies between 
first-order secular theory and N-body simulations, and 
\cite{cuk04} demonstrate that special care must be
given to model correctly the secular motion of irregular 
satellites. \cite{christou97} generalize Laplace-Lagrange secular
theory to second-order in the masses, and apply the result
to the secularly-dominated Uranian satellite system.
Several extrasolar systems exhibit ``hierarchical''
behavior, and ``octupole-level'' secular theory has been 
developed in order to describe analogous systems 
\citep{krymolowski99,fordetal00,lee03}.
The secular phase space of multi-planet systems is complex 
\citep{fejoz02,michtchenko04}, and their accurate
characterization necessitates a comprehensive general analytical
treatment.

Despite these limitations, classical LL theory
remains useful 
\citep{greenberg77,heppenheimer80,chiang02,jietal03,wu03,
zhou03,zamaska04,adams06,namozhou06}
in order to, qualitatively at least, 
describe secular motion.
The formulation of the theory admits some compact analytical results 
\citep{malhotra02, moriwaki04, veras04}
and allows one to circumvent 
singularities present in Lagrange's planetary equations, which
explicitly describe the time evolution of orbital elements of a 
secondary.  Here we extend the planar theory to fourth order, and 
demonstrate the physical consequences of this extension.  In doing so,
we help remove a major assumption (that requiring small eccentricities)
of the standard theory, and address the limits of the theory's
applicability, even in its more general state.

At least $20$ multi-planet extrasolar systems around Solar-type
stars are known to exist \citep{butleretal06}, four of 
which are three-planet systems ($\upsilon$ And, HD 37124, 
HD 69830, GJ 876) and two of which are  
four-planet systems ($\mu$ Ara, 55 Cnc).  The orbital periods for several 
pairs of the planets in these $\approx 20$ systems are in a 
high-order mean motion
commensurability, the effects of which typically dominate
the long-term motion.  However, other pairs of planets are near no
such commensurability, and hence are described by orbital elements which
vary primarily according to secular perturbations.  

Section 2 briefly reviews standard LL secular theory and introduces
some notation.  Section 3 generalizes the theory for two resonant
bodies, and presents some analytical results in specific cases.
The idea of ``apsidal libration'' is discussed in Section 4, and 
criteria for libration are presented.  Section 5 generalizes the 
theory to $N$ resonant bodies, and Section 6
compares the second and fourth-order theories and 
octupole-level theory with numerical 
integrations of several extrasolar planetary systems.
We discuss the implications and conclusions of our model  
in Sections 7 and 8, respectively.

\section{Standard Laplace-Lagrange Secular Theory} 

The fully general disturbing function, $\mathcal{R}_j$, where $j = 1 (2)$
for the outer (inner) secondary, can be written as:

\begin{equation}
\mathcal{R}_j = a_{1}^{-1} \sum_{i=1}^{\infty}
\left[ \sum_{u=1}^{\infty} \mathcal{C}_{j}^{(i,u)} X^{(i,u)} \right]
\cos{\phi}^{(i)}
\label{gendist}
\end{equation}

\noindent{such} that $i$ labels each secular and resonant
argument $\phi$,  $u$ labels each coefficient of a given
argument, $X^{(i,u)}$ is
a function of each secondary's inclination and
eccentricity ($\equiv$ $e_1, e_2$) alone, and $\mathcal{C}_{j}^{(i,u)}$
is a function of each secondary's semimajor axis 
($\equiv$ $a_1, a_2$) alone.

In traditional Laplace-Lagrange theory, only four arguments
are retained, none featuring coupling between eccentricity and
inclination.  As this work is only concerned with the former,
we present only the planar theory in this section.
The two eccentricity arguments retained are
$\phi^{(1)} \equiv 0$ and $\phi^{(2)} \equiv \varpi_1 - \varpi_2$,
where $\varpi$ represents the longitude of pericenter. 
Manipulation yields a disturbing function of the form:

\begin{equation}
\mathcal{R}_j = A_{jj} e_{j}^2 + A_{jk} e_1 e_2 
\cos{\left( \varpi_1 - \varpi_2 \right)}
,
\end{equation}

\noindent{where} the constants in time, $A_{jj}$ and $A_{jk}$, are
functions of each secondary's semimajor axis and mass 
($\equiv m_1, m_2$), such that $k = 1,2$ but $k \ne j$.  This form
of the disturbing function is then inserted into a truncated
version of two of Lagrange's Planetary Equations 
(Brouwer \& Clemence 1961):

\begin{eqnarray}
\dot{e}_{j} &=& -\frac{1}{\mu_{j}^{\frac{1}{2}} a_{j}^{\frac{1}{2}} e_j}
              \frac{\partial \mathcal{R}_j}{\partial \varpi_j}
,
\\
\dot{\varpi_j} &=& \frac{1}{\mu_{j}^{\frac{1}{2}} a_{j}^{\frac{1}{2}} e_j}
              \frac{\partial \mathcal{R}_j}{\partial e_j}
,
\end{eqnarray}

\noindent{where} $\mu_j \equiv \mathcal{G} \left( m_0 + m_j \right)$, 
$\mathcal{G}$ represents the gravitational constant and $m_0$ the central
mass.  In order to remove the eccentricity singularity in the 
denominator, the following auxiliary variables are defined:

\begin{eqnarray}
&h_j = e_j \sin{\varpi_j} \qquad \qquad &h_k = e_k \sin{\varpi_k}
\nonumber\\
&k_j = e_j \cos{\varpi_j} \qquad \qquad &k_k = e_k \cos{\varpi_k}
.
\end{eqnarray}


The disturbing function is then re-expressed in terms of 
$h_j$ and $k_j$ only, and after some algebra, one obtains
the equations of motion:

\begin{equation}
\left(
\begin{array}{c}
{\dot{h}_1} \\
{\dot{h}_2}     
\end{array}
\right)
=
\left(
\begin{array}{cc}
A_{11}  &  A_{12}   \\
A_{21}  &  A_{22}   
\end{array}
\right)
\left(
\begin{array}{c}
k_1  \\
k_2     
\end{array}
\right)
.
\end{equation}

The Laplace-Lagrange secular solution hence reduces to two
eigenvalues and eigenvectors, which may be solved for exactly
given the initial conditions.

\section{Fourth-Order Laplace-Lagrange Secular Theory}

In order to obtain the planar fourth-order solution,
we retain all eccentricity terms up to fourth order in
Eq. (\ref{gendist}).  Therefore, we keep three arguments (with
$\phi^{(3)} \equiv 2\varpi_1 - 2\varpi_2$), some of which
contain more than one relevant term.
The particular functions of $\alpha \equiv a_2/a_1$ which the 
$\mathcal{C}_j$
represent can be found in Appendix B of \cite{murray99}.
Using their $f$ notation, one obtains:

\begin{eqnarray}
\mathcal{R}_1 &=& a_{1}^{-1} \mathcal{G} m_2 
      \bigg[ f_1 + f_2 e_{1}^2 +  f_5 e_{1}^2 e_{2}^2 + f_6 e_{1}^4      
\nonumber\\
      &&+ f_{10} e_1 e_2 \cos{\left( \varpi_1 - \varpi_2 \right)}
\nonumber\\
      &&+ f_{12} e_{1}^3 e_2 \cos{\left( \varpi_1 - \varpi_2 \right)}
\nonumber\\
        &&+ f_{17} e_{1}^2 e_{2}^2 
          \cos{\left[2 \left( \varpi_1 - \varpi_2 \right) \right]}
          \bigg]
,
\end{eqnarray}

\begin{eqnarray}
\mathcal{R}_2 &=& a_{1}^{-1} \mathcal{G} m_1 
      \bigg[ f_1 + f_2 e_{2}^2 +  f_5 e_{1}^2 e_{2}^2 + f_4 e_{2}^4
\nonumber\\
        &&+ f_{10} e_1 e_2 \cos{\left( \varpi_1 - \varpi_2 \right)}
\nonumber\\
        &&+ f_{11} e_{1} e_{2}^3 \cos{\left( \varpi_1 - \varpi_2 \right)}
\nonumber\\
        &&+ f_{17} e_{1}^2 e_{2}^2 
          \cos{\left[ 2 \left( \varpi_1 - \varpi_2 \right) \right]}
          \bigg]
.
\end{eqnarray}

We also have:

\begin{mathletters}
\begin{eqnarray}
&\dot{h}_j = &\frac{h_j}{e_j} \dot{e}_j + k_j \dot{\varpi}_j
,
\label{hke1}
\\
&\dot{k}_j = &\frac{k_j}{e_j} \dot{e}_j - h_j \dot{\varpi}_j
,
\label{hke2}
\\
&\dot{e}_j = &-a_{j}^{-\frac{1}{2}} \mu_{j}^{-\frac{1}{2}}
              \frac{\sqrt{1 - e_{j}^2}}{e_{j}}
              \frac{\partial \mathcal{R}_j}{\partial \varpi_j}
,
\label{hke3}
\\
&\dot{\varpi}_j = &a_{j}^{-\frac{1}{2}} \mu_{j}^{-\frac{1}{2}}
              \frac{\sqrt{1 - e_{j}^2}}{e_{j}}
              \frac{\partial \mathcal{R}_j}{\partial e_j}
.
\label{hke4}
\end{eqnarray}
\end{mathletters}

\noindent{In} the traditional Laplace-Lagrange solution, the square
root in the numerator of Eqs. (\ref{hke3})-(\ref{hke4}) 
is expanded only to zeroth-order.
The traditional LL theory demonstrates that the partial derivatives should be
expressed in terms of $h_j$, $h_k$, $k_j$ and $k_k$.  We can still
express the more general disturbing function in terms of these variables
only, while not retaining any eccentricity terms in the denominator, by
using the following forms:

\begin{eqnarray}
e_1 \frac{\partial \mathcal{R}_1}{\partial e_1} =&&
a_{1}^{-1} \mathcal{G} m_2 
      \bigg[ 2 f_2 \left( h_{1}^2 + k_{1}^2 \right)
\nonumber\\
      &&+  2 f_5 \left( h_{1}^2 + k_{1}^2 \right) \left( h_{2}^2 + k_{2}^2 \right)
\nonumber\\
      &&+ 4 f_6 \left( h_{1}^2 + k_{1}^2 \right)^2 
        + f_{10} \left( h_1 h_2 + k_1 k_2 \right)
\nonumber\\
      &&+ 3 f_{12} \left( h_{1}^2 + k_{1}^2 \right) \left( h_1 h_2 + k_1 k_2 \right)
\nonumber\\
      &&+ 2 f_{17} 
        \bigg\lbrace  
        2 \left( h_1 h_2 + k_1 k_2 \right)^2 -   
\nonumber\\
      &&\left( h_{1}^2 + k_{1}^2 \right) \left( h_{2}^2 + k_{2}^2 \right)
        \bigg\rbrace
      \bigg]
,
\end{eqnarray}

\begin{eqnarray}
e_2 \frac{\partial \mathcal{R}_2}{\partial e_2} =&& 
a_{1}^{-1} \mathcal{G} m_1 
      \bigg[ 2 f_2 \left( h_{2}^2 + k_{2}^2 \right)
\nonumber\\
      &&+  2 f_5 \left( h_{1}^2 + k_{1}^2 \right) \left( h_{2}^2 + k_{2}^2 \right)
\nonumber\\
      &&+ 4 f_4 \left( h_{2}^2 + k_{2}^2 \right)^2 
        + f_{10} \left( h_1 h_2 + k_1 k_2 \right)
\nonumber\\
      &&+ 3 f_{11} \left( h_{2}^2 + k_{2}^2 \right) \left( h_1 h_2 + k_1 k_2 \right)
\nonumber\\
      &&+ 2 f_{17} 
        \bigg\lbrace  
        2 \left( h_1 h_2 + k_1 k_2 \right)^2 - 
\nonumber\\  
      &&\left( h_{1}^2 + k_{1}^2 \right) \left( h_{2}^2 + k_{2}^2 \right)
        \bigg\rbrace
      \bigg]
,
\end{eqnarray}

\begin{eqnarray}
\frac{\partial \mathcal{R}_1}{\partial \varpi_1} =&& 
a_{1}^{-1} \mathcal{G} m_2 
      \bigg[  - f_{10} \left( h_1 k_2 - h_2 k_1 \right) 
\nonumber\\
      &&- f_{12} \left( h_1 k_2 - h_2 k_1 \right) \left( h_{1}^2 + k_{1}^2 \right)
\nonumber\\
      &&- 4 f_{17} 
          \left( h_1 k_2 - h_2 k_1 \right) \left( h_1 h_2 + k_1 k_2 \right)
          \bigg]
,
\end{eqnarray}

\begin{eqnarray}
\frac{\partial \mathcal{R}_2}{\partial \varpi_2} =&& 
a_{1}^{-1} \mathcal{G} m_1 
      \bigg[  f_{10} \left( h_1 k_2 - h_2 k_1 \right) 
\nonumber\\
      &&+ f_{11} \left( h_1 k_2 - h_2 k_1 \right) \left( h_{2}^2 + k_{2}^2 \right)
\nonumber\\
      &&+ 4 f_{17} 
          \left( h_1 k_2 - h_2 k_1 \right) \left( h_1 h_2 + k_1 k_2 \right)
          \bigg]
\label{lastpar}
.
\end{eqnarray}

Equations (\ref{hke1})-(\ref{lastpar})
may be manipulated in order
to generate the 4th-order analog of the LL theory's
differential equations of motion:

\begin{eqnarray}
\dot{h}_j  &&=  C_{a}^{(j)}
   \bigg[
    C_1 k_j 
  + C_2 k_k 
  + C_{3}^{(j)} k_j h_{j}^2
\nonumber\\
  &&+ C_4 k_j h_{k}^2
  + C_5 k_j k_{k}^2
  + C_{6}^{(j)} k_{j}^2 k_k
  + C_{7}^{(j)} k_k h_{j}^2 
\label{eqmo1}
\nonumber\\
 &&+ C_{8}^{(j)} k_j h_j h_k
  + C_9 k_k h_j h_k  
  + C_{3}^{(j)} k_{j}^3 
   \bigg]
,
\\
&&
\nonumber\\
\dot{k}_j  &&=  - C_{a}^{(j)}
   \bigg[
    C_1 h_j 
  + C_2 h_k 
  + C_{3}^{(j)} h_j k_{j}^2
\nonumber\\
  &&+ C_4 h_j k_{k}^2
  + C_5 h_j h_{k}^2
  + C_{6}^{(j)} h_{j}^2 h_k
  + C_{7}^{(j)} h_k k_{j}^2 
\nonumber\\
 &&+ C_{8}^{(j)} h_j k_j k_k
  + C_9 h_k k_j k_k  
  + C_{3}^{(j)} h_{j}^3 
   \bigg]
,
\label{eqmo}
\end{eqnarray}

\noindent{where} all the $C_i$, $i = 1...9$, are constant functions
of $\alpha$, and the scaling factors $C_{a}^{(j)}$ are constant functions
of both secondary masses and semimajor axes.  These constant values are 
provided in Appendix A.  This form of the equations illustrates well the 
dependence of all four variables on 
one another, the symmetry of those variables, and the asymmetry in some of 
the constants.  By retaining only the first two terms in the brackets of 
these equations, one recovers traditional Laplace-Lagrange secular theory.

Equations (\ref{eqmo1})-(\ref{eqmo}) are a set of four first-order 
coupled highly nonlinear differential equations, and in general 
need to be solved numerically.  However, in the special case
where one secondary on a circular orbit is very massive compared 
to the other secondary, we can obtain an analytic result.  
Assuming $e_k = 0$ for all time, we need only to solve:

\begin{eqnarray}
&\dot{h}_j 
    &= C_{a}^{(j)} k_j \left[ C_1 + C_{3}^{(j)} \left( h_{j}^2 + k_{j}^2 \right) \right] 
\\
&\dot{k}_j 
  &= - C_{a}^{(j)} h_j \left[ C_1 + C_{3}^{(j)} \left( h_{j}^2 + k_{j}^2 \right) \right]
\label{spcase}
.
\end{eqnarray}

This set of equations is also nonlinear; however, the presence of the squared
quantities inside the parenthesis suggests a sinusoidal solution.  
Assuming body $j$'s initial eccentricity equals $e_0$, the solution is:

\begin{eqnarray}
&h_j(t) &= e_0 \sin{
\left[ C_{a}^{(j)} \left( C_1 + e_{0}^2 C_{3}^{(j)} \right) t \right]} 
\label{rate1}
\\
&k_j(t) &= e_0 \cos{
\left[C_{a}^{(j)} \left( C_1 + e_{0}^2 C_{3}^{(j)} \right) t \right]}
\label{rate2}
.
\end{eqnarray}

\noindent{and} therefore body $j$'s longitude of pericenter varies
linearly with a constant of proportionality equal to 
$C_{a}^{(j)} \left( C_1 + e_{0}^2 C_{3}^{(j)} \right)$.
In the traditional Laplace-Lagrange solution, this rate is
independent on the initial eccentricity.  We can 
analyze this rate further by expressing it in terms of orbital
elements (details of the derivation are found in Appendix B):

\begin{eqnarray}
&C_1 + e_{0}^2 C_{3}^{(1)} &\approx  
\frac{3}{4} \alpha^2 \left( 1 + 2 e_{0}^2 \right)
\nonumber\\
&&+
\frac{45}{32} \alpha^4 \left( 1 + \frac{19}{4} e_{0}^2 \right) + \mathcal{O} \left( \alpha^6 \right)
\\
&C_1 + e_{0}^2 C_{3}^{(2)} &\approx 
\frac{3}{4} \alpha^2 \left( 1 - \frac{1}{2} e_{0}^2 \right)
\nonumber\\
&&+
\frac{45}{32} \alpha^4 \left( 1 + \frac{1}{4} e_{0}^2 \right) + \mathcal{O} \left( \alpha^6 \right)
.
\label{rate2}
\end{eqnarray}

The above equations demonstrate that the secular perturbations between
a massive body on a circular orbit and a much less massive 
eccentric {\it exterior} body will cause the latter's pericenter rate 
to increase with a greater initial eccentricity.  Alternatively,
a less massive eccentric body {\it interior} to a massive object
will precess at a slower rate given a greater initial eccentricity. 
Such dependencies are not present in the traditional Laplace-Lagrange
secular theory, which predicts (under the assumption of a massive
central body), the rates to be equal.

Furthermore, by making use of the critical planetary separation 
needed to achieve 
Hill Stability \citep{gladman93,veras04}, we can derive 
an upper bound for the rate of change at pericenter or precession
of a perturbed object exterior
to a massive planet (orbiting a much more massive star), 
in units of radians/year:

\begin{equation}
\left( \dot{\varpi}_{1} \right)_{\rm max} 
\approx
\frac
{
3 m_1 a_{2}^{\frac{3}{2}}
}
{
4  \left(  
a_2 + 2 \cdot 3^{\frac{1}{6}} m_{1}^{\frac{1}{3}}
\right)^3
}
,
\label{hill}
\end{equation}

\noindent{where} $m_1$ is measured in solar masses and $a_2$ is measured
in AUs.  For $m_1 = m_{Jup}$ and $a_2 = 1$ AU, 
$\left( \dot{\varpi}_{1} \right)_{\rm max}$ gives a maximum rate corresponding
to a $\approx 16,000$ yr period.

\section{Apsidal Libration Amplitudes}

In traditional LL theory, one nonzero argument in the 
disturbing function exists whose time derivative 
is equal to $\left( \dot{\varpi_1} - \dot{\varpi_2} \right)$, a quantity 
known as the ``apsidal rate''.  When this rate librates, 
rather than circulates,
it is sometimes claimed that the system is in ``apsidal resonance'', 
which has been the subject of much
study \citep{breiter99,malhotra02,chiang02,beauge03,zhou03}.
As observed by 
\cite{barnes06}, the definition of ``resonance" has been 
multiply-defined in the literature.  Figure 9.3 of \cite{morby02}
demonstrates that in fact libration and resonant regions of 
phase space may be distinct and/or independent.
We don't expound on this matter here, but instead observe 
that libration and resonance are typically
closely linked, and that an account of the former often
aids in describing the latter.

Librational motion requires
that the sign of $\phi^{(2)}$ ($ = \varpi_1 - \varpi_2$) change.  The amplitude
of libration, often used to indicate how ``deep'' the system is in resonance,
may be found by evaluating the value of $\phi^{(2)}$ satisfying
$\dot{\phi}^{(2)} = 0$.  The resulting amplitude, $\phi^{(2)}_{\rm max}$, is:

\begin{eqnarray}
&&\cos{\left( \phi^{(2)}_{\rm max} \right)}
= 
\nonumber\\
&&\frac{C_1}{C_2} 
\left[
\frac{
C_{a}^{(2)} e_{2}^2 - C_{a}^{(1)} e_{1}^2
}
{
\left( C_{a}^{(1)} - C_{a}^{(2)} \right) e_1 e_2 
}
\right]
.
\label{crit2}
\end{eqnarray}

Equation (\ref{crit2}) may be compared directly with Eqs. (16)-(21) of 
\cite{barnes06}.  The derivation of Eq. (\ref{crit2}) 
suggests that no apsidal libration occurs
if either body's orbit is circularized, and even if both objects harbor
eccentric orbits, the absolute value of the right-hand-side of the equation 
must be less than unity.  These represent the conditions for apsidal 
libration to occur according to standard LL theory.

For equal mass bodies, 

\begin{equation}
\cos{\left( \phi^{(2)}_{\rm max} \right)}
= \frac{C_1}{C_2} 
\left[
\frac{
e_{2}^2 - \sqrt{\alpha} e_{1}^2
}
{
\left( 1 - \sqrt{\alpha} \right) e_1 e_2 
}
\right]
,
\end{equation}

\noindent{whereas} for equally eccentric bodies, 
$\phi^{(2)}_{\rm max} \approx \cos^{-1} \left( C_1/C_2 \right)$.
One must take care to not approximate the ratio $ C_1/C_2$
by low-order powers of $\alpha$, as convergence can be
slow, requiring several tens of terms of $\alpha$.  A numerical
sampling of phase space indicates that this ratio is less than
unity when $\alpha \ge 0.409$.  This result is independent
of the masses or eccentricities, assuming that the eccentricities of both
bodies are equal.  Physically, for separations
greater than those given by this criterion, $\phi^{(2)}$ must
circulate, as in this case, the bodies are too far from each
other to be ``in resonance''.  On the other extreme, if the bodies
are too close to each other, the system will become unstable due
to the saturation of mean motion resonances.  Wisdom (1980) provides
an approximate criterion for this critical distance, which is proportional
to the 2/7ths power of the primary-secondary mass ratio.  If $\alpha$
lies between these two extremes, {\it and} is sufficiently far
from any sufficiently strong mean motion resonance, then standard LL theory
predicts apsidal libration will occur.

In the fourth-order theory, many more terms are included in the 
disturbing function.
However, the relation $\phi^{(3)} = 2 \phi^{(2)}$ allows us to 
derive the apsidal librational amplitude:

\begin{eqnarray}
&&\cos{\left( \phi^{(2)}_{\rm max} \right)}
=
\nonumber\\ 
&&\frac{
C_{a}^{(2)} \left( C_2 + C_{6}^{(2)} e_{2}^2 \right) - 
C_{a}^{(1)} \left( C_2 + C_{6}^{(1)} e_{1}^2 \right)
\pm
\sqrt{D}
}
{
2 C_{9} e_1 e_2 \left( C_{a}^{(1)} - C_{a}^{(2)}  \right)
}
\label{crit4}
\end{eqnarray}

\noindent{where}

\begin{eqnarray}
&D &= \left[
C_{a}^{(2)} \left( C_2 + C_{6}^{(2)} e_{2}^2 \right)
-
C_{a}^{(1)} \left( C_2 + C_{6}^{(1)} e_{1}^2 \right)
\right]^2 +
\nonumber\\
&&
4 C_9 \left( C_{a}^{(1)} - C_{a}^{(2)} \right)
\bigg[  
e_{2}^2 C_{a}^{(2)} 
\left( C_1 + C_{3}^{(2)} e_{2}^2 + C_4 e_{1}^2 \right)
\nonumber\\
&&-
e_{1}^2 C_{a}^{(1)} 
\left( C_1 + C_{3}^{(1)} e_{1}^2 + C_4 e_{2}^2 \right)
\bigg]
\label{det}
\end{eqnarray}

The different solution branches manifest
themselves in a myriad of ways, and are multiply-dependent on
the eccentricities and semimajor axes.  Recall that
in the limit of one planet of negligible mass on an eccentric orbit,
its eccentricity remains constant, while its longitude of pericenter
precesses at a constant rate, thereby preventing apsidal libration.
Therefore, Eq. (\ref{crit4}) only applies in the case of two nonzero 
eccentricities.  Yet, the equation is well-behaved as a circular 
orbit is approached.

\section{N-body formulation}

Extrasolar planetary systems with at least three planets are
being discovered in increasing numbers, and the Solar System
provides an example of a system in which several terrestrial
and giant planets may coexist in a configuration stable over
timescales of at least several Myr.  A secular
theory which could accurately describe such systems would be useful.
We here present the fourth-order LL theory for $N$ bodies.
The disturbing function for secondary $j$, $j = 1...N$, is:

\begin{eqnarray}
&\mathcal{R}_j&=\sum_{k=1, k \ne j}^N  a_{l}^{-1} \mathcal{G} m_k 
      \bigg\lbrace f_{1}^{(lp)} + f_{2}^{(lp)} e_{j}^2 +  
             f_{5}^{(lp)} e_{j}^2 e_{k}^2 
\nonumber\\
          &&+ 
             \left[ 
           \delta \left( j-p \right) f_{4}^{(lp)} + 
           \delta \left( k-p \right) f_{6}^{(lp)} 
             \right]
\nonumber\\
      &&\cdot e_{j}^4 + f_{10}^{(lp)} e_j e_k \cos{\left( \varpi_l - \varpi_p \right)}
\nonumber\\
      &&+ 
             \left[ 
           \delta \left( j-p \right) f_{11}^{(lp)} + 
           \delta \left( k-p \right) f_{12}^{(lp)} 
             \right] e_{j}^3 e_k
\nonumber\\
       &&\cdot \cos{\left( \varpi_l - \varpi_p \right)}
         + f_{17}^{(lp)} e_{j}^2 e_{k}^2 
          \cos{\left[2  \left(\varpi_l - \varpi_p \right) \right]}
          \bigg\rbrace
\label{general}
\end{eqnarray}

\noindent{where} $l \equiv \min{\left( j,k \right)}$, $p \equiv \max{\left( j,k \right)}$,
and $f_{i}^{(lp)} \equiv f_{i} \left( a_p/a_l \right)$.  
After following the same procedure used to derive Eq. (\ref{eqmo}),
one obtains,

\begin{eqnarray}
\dot{h}_j  &&=  
\sum_{k=1, k \ne j}^N
C_{a}^{(lp)}
   \bigg[
    C_{1}^{(lp)} k_j 
  + C_{2}^{(lp)} k_k 
  + C_{3}^{(lp)} k_j h_{j}^2
\nonumber\\
  &&+ C_{4}^{(lp)} k_j h_{k}^2
  + C_{5}^{(lp)} k_j k_{k}^2
\label{eqmo2a}
\nonumber\\
 &&+ C_{6}^{(lp)} k_{j}^2 k_k
  + C_{7}^{(lp)} k_k h_{j}^2 
  + C_{8}^{(lp)} k_j h_j h_k
\nonumber\\
  &&+ C_{9}^{(lp)} k_k h_j h_k  
  + C_{3}^{(lp)} k_{j}^3 
   \bigg]
,
\\
&
\nonumber\\
\dot{k}_j  &&=  -
\sum_{k=1, k \ne j}^N
    C_{a}^{(lp)}  \bigg[
    C_{1}^{(lp)} h_j 
  + C_{2}^{(lp)} h_k 
  + C_{3}^{(lp)} h_j k_{j}^2
\nonumber\\
  &&+ C_{4}^{(lp)} h_j k_{k}^2
  + C_{5}^{(lp)} h_j h_{k}^2
\nonumber\\
 &&+ C_{6}^{(lp)} h_{j}^2 h_k
  + C_{7}^{(lp)} h_k k_{j}^2 
  + C_{8}^{(lp)} h_j k_j k_k
\nonumber\\
  &&+ C_{9}^{(lp)} h_k k_j k_k  
  + C_{3}^{(lp)} h_{j}^3 
   \bigg]
\label{eqmo2}
,
\end{eqnarray}

\noindent{with} new constants labeled $C_{i}^{(lp)}$ whose explicit
form is provided in Appendix A.
In order to acquire a qualitative perspective on the
secular motion of an eccentric terrestrial planet in
the midst of several giant planets on circular orbits,
assume all bodies except body $j$ are massive with 
zero eccentricities.  Then,

\begin{eqnarray}
h_j(t) &= e_0 \sin{
\left[ 
\sum_{k=1, k \ne j}^N
C_{a}^{(lp)} \left( C_{1}^{(lp)} + e_{0}^2 C_{3}^{(lp)} \right)
t \right]} 
,
\nonumber\\
k_j(t) &= e_0 \cos{
\left[
\sum_{k=1, k \ne j}^N
C_{a}^{(lp)} \left( C_{1}^{(lp)} + e_{0}^2 C_{3}^{(lp)} \right)
t \right]}
\end{eqnarray}

The criteria for libration can be evaluated using the N-body
formulation to determine, for example, 
if the apsidal libration of terrestrial planets interior
to a giant planet is greater than if the terrestrial planets
were exterior to the giant planet.
Consider a pair of terrestrial masses in the same system with
a giant planet, and denote $\mathcal{M}$ as the terrestrial planet/giant planet 
mass ratio.  
The giant planet may reside in between ($\equiv$ Case 1), exterior
to ($\equiv$ Case 2), or interior to ($\equiv$ Case 3) the terrestrial 
planets.  In all cases, the planets are labeled such that 
$a_1 > a_2 > a_3$, and the
arguments $\phi$ such that $\phi_{lp} \equiv 
\left( \varpi_l - \varpi_p \right)_{\rm max}$.  According to the second-order
theory, and assuming the giant planet is much more massive than each 
terrestrial planet, the resulting amplitudes of the librating angles 
may be expressed as:

\begin{eqnarray}
&&{\rm Case}\; 1:
\nonumber\\
&&\cos{\phi_{13}} = \left( \frac{2}{\mathcal{M}} \right)
              \frac{f_{2}^{(23)} e_{3}^2
                -f_{2}^{(12)} \sqrt{\alpha_{13}} \alpha_{12} e_{1}^2}
                {f_{10}^{(13)} \alpha_{12} 
                \left( \sqrt{\alpha_{13}} - 1 \right) e_1 e_3},
\label{threephi1}
\\
&&{\rm Case}\; 2:
\nonumber\\
&&\cos{\phi_{23}} = \left( \frac{2}{\mathcal{M}} \right)
              \frac{f_{2}^{(13)} \alpha_{12}  e_{3}^2
                -f_{2}^{(12)} \sqrt{\alpha_{23}} e_{2}^2 }
                {f_{10}^{(13)} 
                \left( \sqrt{\alpha_{23}} - 1  \right) e_2 e_3},
\label{threephi2}
\\
&&{\rm Case}\; 3:
\nonumber\\
&&\cos{\phi_{12}} = \left( \frac{2}{\mathcal{M}} \right)
              \frac{f_{2}^{(23)} e_{2}^2
                -f_{2}^{(13)} \sqrt{\alpha_{12}} \alpha_{12} e_{1}^2 }
                {f_{10}^{(12)} \alpha_{12}  
                \left( \sqrt{\alpha_{12}} - 1 \right) e_1 e_2},
\label{threephi3}
\end{eqnarray}

\noindent{where} we have defined $\alpha_{lp} \equiv a_p/a_l$.
We can use Eqs. (\ref{threephi1})-(\ref{threephi3}) to 
sample regions of semimajor
axis and eccentricity phase space in which apsidal librational 
may occur.  
For either an interior or exterior pair of terrestrial 
planets, libration may only occur if they are close to each other 
but far from the giant planet - in effect, when the terrestrial planets
represent an isolated system.  In no instance does
a giant planet embedded in between the terrestrial planets allow
for apsidal libration.

Although such formulas may apply in Solar System analogs, with
giant planets on near-circular orbits, known extrasolar systems
commonly harbor giant planets on moderately or highly-eccentric
orbits.  The origin of these eccentricities remains unclear,
but may be reproduced based on gravitational scattering among
multiple massive planets \citep{rasio96,
lin97,levilissdunc98,marzari02,ford03a, ford05, verasarm06} or 
disk-planet interactions 
\citep{artymowicz91,papaloizou01,ogilvie03,goldreich03}.
If we assume that a pair of terrestrial planets coexists with an
{\it eccentric} giant planet, then the angles
$\left( \varpi_1 - \varpi_3 \right)$, 
$\left( \varpi_2 - \varpi_3 \right)$
and $\left( \varpi_1 - \varpi_2 \right)$ are all involved in the criteria
for libration.  The maximum librational angles may be manipulated
into the following form:

\begin{equation}
\cos{\phi_{lp}} = \frac{ - Q^{(lp)} R^{(lp)}  \pm S^{(lp)} 
                     \sqrt{Q^{{(lp)}^2} + S^{{(lp)}^2} - R^{{(lp)}^2}}  }
                    {Q^{{(lp)}^2} + S^{{(lp)}^2}}
,
\label{RQS}
\end{equation}

\noindent{where} formulas for $Q^{(lp)}$, $R^{(lp)}$, and $S^{(lp)}$,
which are all functions of $\alpha_{13}, \alpha_{23}, \alpha_{12}, e_1,
e_2, e_3$, and one additional angle, are provided in Appendix C.
This additional angle is equal to $\left( \varpi_1 - \varpi_2 \right)$ for 
$p=3$ and $ \left(\varpi_1 - \varpi_3 \right)$  for $p=2$.
The free parameter introduced by this additional angle is multiplied
by the eccentricity of the giant planet, which is assumed to be zero in 
Eqs. (\ref{threephi1})-(\ref{threephi3}).  The free parameter has a marked
effect on the subsequent phase space analysis, and such an analysis
may encompass an entire study. 

\begin{figure}
\centerline{\psfig{figure=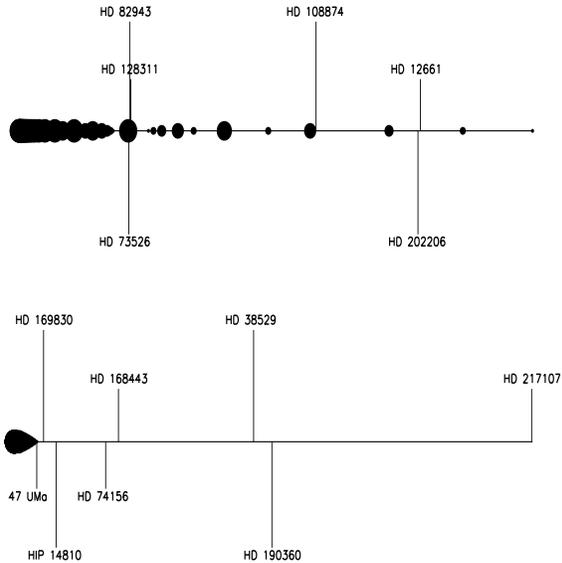,width=\columnwidth,height=3.5truein}}
\figcaption{
Effective distance from mean motion resonance for all known pairs 
of planets in two-planet multiplanet systems, measured in 
semimajor axis ratio.  Locations of 1st, 2nd, 3rd, 4th,
5th and 6th order mean motion resonances are given by dots of 
decreasing size.  The
lower horizontal line is on a broader scale than the top line.
\label{plotone}}
\end{figure}

\begin{figure}
\centerline{\psfig{figure=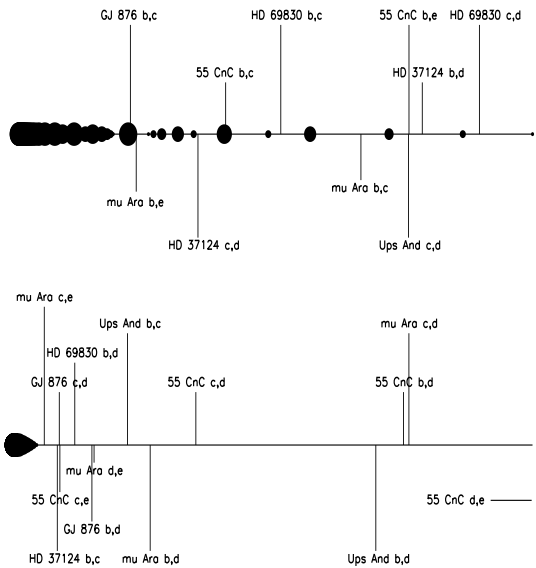,width=\columnwidth,height=3.5truein}}
\figcaption{
Effective distance from mean motion resonance for all known pairs 
of planets in multiplanet systems with more than two planets, measured in 
semimajor axis ratio.  Locations of 1st, 2nd, 3rd, 4th,
5th and 6th order mean motion resonances are given by dots of 
decreasing size.  The
lower horizontal line is on a broader scale than the top line.
\label{plottwo}}
\end{figure}

\section{Application to known exosystems}

\subsection{Overview}

Extrasolar planet data is updated constantly, and the resulting orbital
solutions are bound to change with the passage of time.  Nevertheless,
in many cases the errors on currently known orbital parameters allow one
to achieve a representative qualitative description of the motion,
and often a highly accurate quantitative solution as well. 
In order to help determine which extrasolar planets are dominated by
secular perturbations instead of mean motion resonances (MMRs), we compute
the ``effective distance'' away from a strong MMR for each pair of planets.  
The ``nominal'' (approximate) 
locations of MMRs are given by $[(p+n)/p]^{-(2/3)}$, where $n$
represents the order of the resonance and $p$ is an integer.  Therefore,
a scale-free method of representing this ``effective distance'' from resonance
is to use this quantity as the metric \citep{champenois99}.
Figures \ref{plotone}-\ref{plottwo} provide a summary of our 
current (as of Jan. 1, 2007) 
understanding of all known multiple-planet extrasolar systems. 
A schematic of the effective distance from resonance is provided 
for all two-planet systems (Fig. \ref{plotone}) and all known 
systems with more than two planets (Fig. \ref{plottwo}).
Table 1 lists the relevant orbital parameters for the systems simulated
in this study.  We deem these systems to be sufficiently far from any 
MMR, and hence ``secular''.  All orbital parameters are taken 
from Jean Schneider's Extrasolar Planets Encyclopedia\footnote{at http://vo.obspm.fr/exoplanetes/encyclo/encycl.html}, and for all data we take $\sin{i}$ to be unity.
All N-body numerical integrations are carried out
with the Bulirsch-Stoer HNbody integrator (K.P. Rauch \& D. P. Hamilton
2007, in preparation).

\begin{table}
\tablecaption{?-PLANET EXTRASOLAR SYSTEMS}
\begin{center}
\begin{tabular}{c  c  c  c  c  c}
Planet     & $m$ ($m_{Jup}$) & $m_{\ast}$ ($m_{\odot}$) & $a$ (AU)& $e$ & $\varpi$ (deg)  \\
\hline\hline
HD 12661 b & 2.30 & & 0.83 & 0.35 & 291.73 \\  \cline{1-2}\cline{4-6}
HD 12661 c & 1.57 & \raisebox{1.5ex}[0pt]{1.07} & 2.56 & 0.2 & 162.4  \\ \hline

HD 38529 b & 0.78 & & 0.129 & 0.29 & 87.7  \\ \cline{1-2}\cline{4-6}
HD 38529 c & 12.7 & \raisebox{1.5ex}[0pt]{1.39} & 3.68 & 0.36 & 14.7 \\ \hline

HD 168443 b & 8.02 & & 0.30 & 0.5286 & 172.87 \\ \cline{1-2}\cline{4-6}
HD 168443 c \tablenotemark{a}
\tablenotetext{a}{Considered to be a brown dwarf by some due to its mass}
& 18.1 & \raisebox{1.5ex}[0pt]{0.96}& 3.91 & 0.2125 & 65.07   \\ \hline

HD 108874 b & 1.36 & & 1.051 & 0.07 & 248.4  \\ \cline{1-2}\cline{4-6}
HD 108874 c & 1.018 & \raisebox{1.5ex}[0pt]{1.00} & 2.68 & 0.25 & 17.3 \\ \hline
$\upsilon$ And b  
& 0.690 & & 0.0590 & 0.029 & 46.0 \\ 
\cline{1-2}\cline{4-6}
$\upsilon$ And c
& 1.98 & 1.30 & 0.830 & 0.254 & 232.4  \\ 
\cline{1-2}\cline{4-6}
$\upsilon$ And d 
& 3.95 & & 2.51 & 0.242 & 258.5   \\ \hline\hline
\end{tabular}
\caption{  \small{Summary of masses, semimajor axes, 
eccentricities and longitude of pericenters for the 
multi-planet extrasolar systems studied here.  These parameters 
were taken from Jean Schneider's Extrasolar Planets 
Encyclopedia with $\sin{i}$ assumed to be unity.} }
\end{center}
\end{table}

We study only those planets which satisfy a particular constraint
regarding Kaula's (1961; 1962) expansion of the disturbing function, 
which is based on Laplace 
coefficients.  \cite{sundman16} determined that in the planar
form of this expansion,
Laplace coefficients are sure to 
converge only if the following criterion is satisfied 
\citep{ferrazmello94,sid94}:

\begin{equation}
a_2 H(e_2) < a_1 h(e_1)
,
\end{equation}

\noindent{where}

\begin{eqnarray}
H(q) &&= \sqrt{1 + q^2} \cosh{w} + q + \sinh{w}
,
\\
h(q) &&= \sqrt{1 + q^2} \cosh{w} - q - \sinh{w}
,
\end{eqnarray}

\noindent{such} that $w$ is implicitly defined as $w = q \cosh{w}$.
We applied this test to each pair of planets in the same system;
failing this test implies that the use of LL theory, or any other theory
relying on Laplace coefficients, is suspect.  Therefore, we only
consider pairs of planets which pass this test.

\subsection{Systems with two planets}

\subsubsection{HD 12661}

\cite{lee03} use octupole-level secular theory and 
\cite{rodriguez05} use a sixth-order expansion
of Kaula's disturbing function  in order to study the
planets orbiting HD 12661, 
and conclude that the 
system is strongly dominated by secular motions.  However, older work 
indicated that the system might be trapped in a $6$:$1$ 
\citep{gozdziewskima03} resonance or is on the border of a $11$:$2$
\citep{gozdziewski03} MMR.

\begin{figure}
\centerline{\psfig{figure=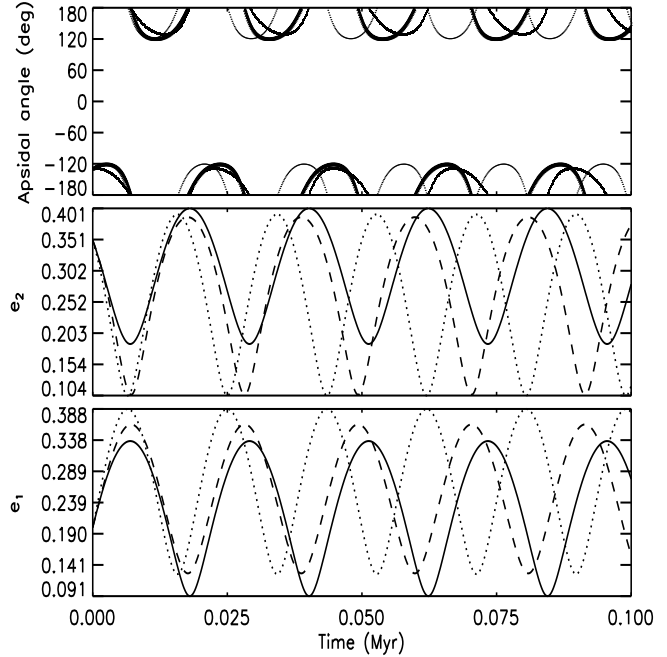,width=\columnwidth,height=3.5truein}}
\figcaption{
Eccentricity and apsidal angle variation according to  
the traditional LL theory (solid lines in the lower two panels and leading
dark loops in the uppermost panel), the fourth-order LL theory 
(dotted lines in the lower two panels and light loops in the uppermost panel),
and octupole-level secular perturbation theory 
(dashed lines in the lower two panels and 
trailing dark loops in the uppermost panel) for the HD 12661 system.
\label{plotthree}}
\end{figure}

\begin{figure}
\centerline{\psfig{figure=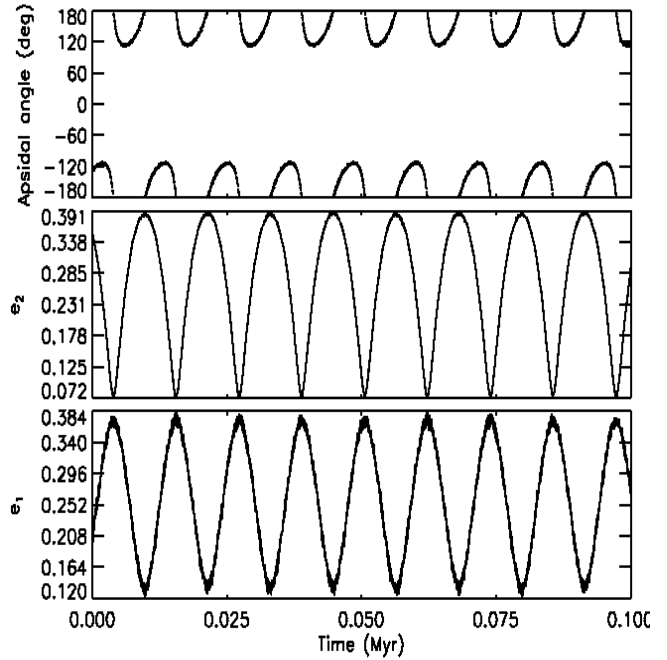,width=\columnwidth,height=3.5truein}}
\figcaption{
Eccentricity and apsidal angle variation from an N-body simulation 
for the HD 12661 system.
\label{plotfour}}
\end{figure}

We seek to determine how well the traditional and generalized versions
of LL theory reproduce motion in HD 12661, and further compare these methods
with octupole-level secular theory.
Figs. \ref{plotthree}-\ref{plotfour} explore the dynamical state of 
the system, and demonstrate that none of the theories
employed here model the true evolution well. 
Considering first the variation of the eccentricity 
of $e_1$, the fourth-order LL theory matches best the
secular periodicity, eccentricity amplitude, and eccentricity values
of the profile exhibited by the true evolution.  Octupole 
theory matches fourth-order LL
theory in their predictions for the lower bounds of the eccentricity
of both planets, and both theories represent a significant 
improvement over traditional LL theory with regard to eccentricity
amplitude.  All theories and the actual dynamical evolution
exhibit apsidal libration with a $\approx 60^{\circ}$ amplitude.
Overall, the profiles and periodicities of the apsidal libration
for both the fourth-order LL theory and octupole theory
improve upon the traditional theory, with the frequency from
the fourth-order LL theory mirroring the actual evolution
best but still not well.

\subsubsection{HD 168443}

\begin{figure}
\centerline{\psfig{figure=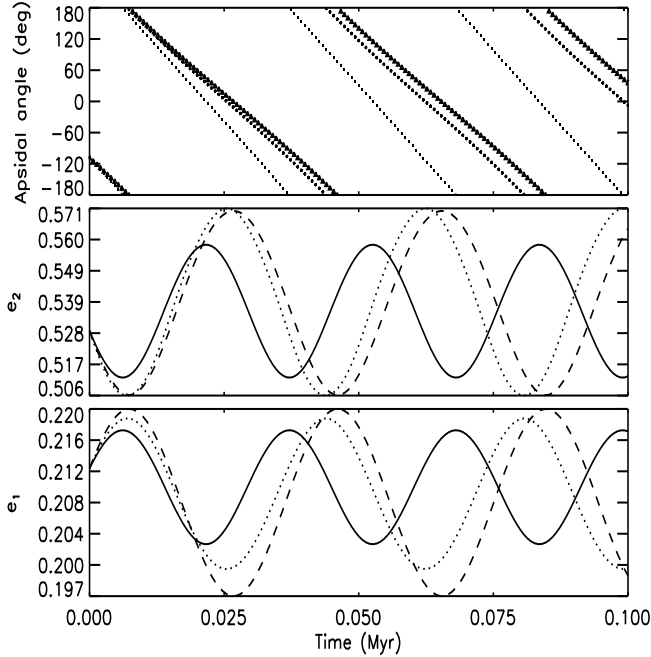,width=\columnwidth,height=3.5truein}}
\figcaption{
Eccentricity and apsidal angle variation according to    
the traditional LL theory (solid lines in the lower two panels and light
squares in the uppermost panel), the fourth-order LL theory 
(dotted lines in the lower two panels and dots in the uppermost panel),
and octupole-level secular perturbation theory 
(dashed lines in the lower two panels and thick triangles
in the uppermost panel) for the HD 168443 system.
\label{plotfive}}
\end{figure}

\begin{figure}
\centerline{\psfig{figure=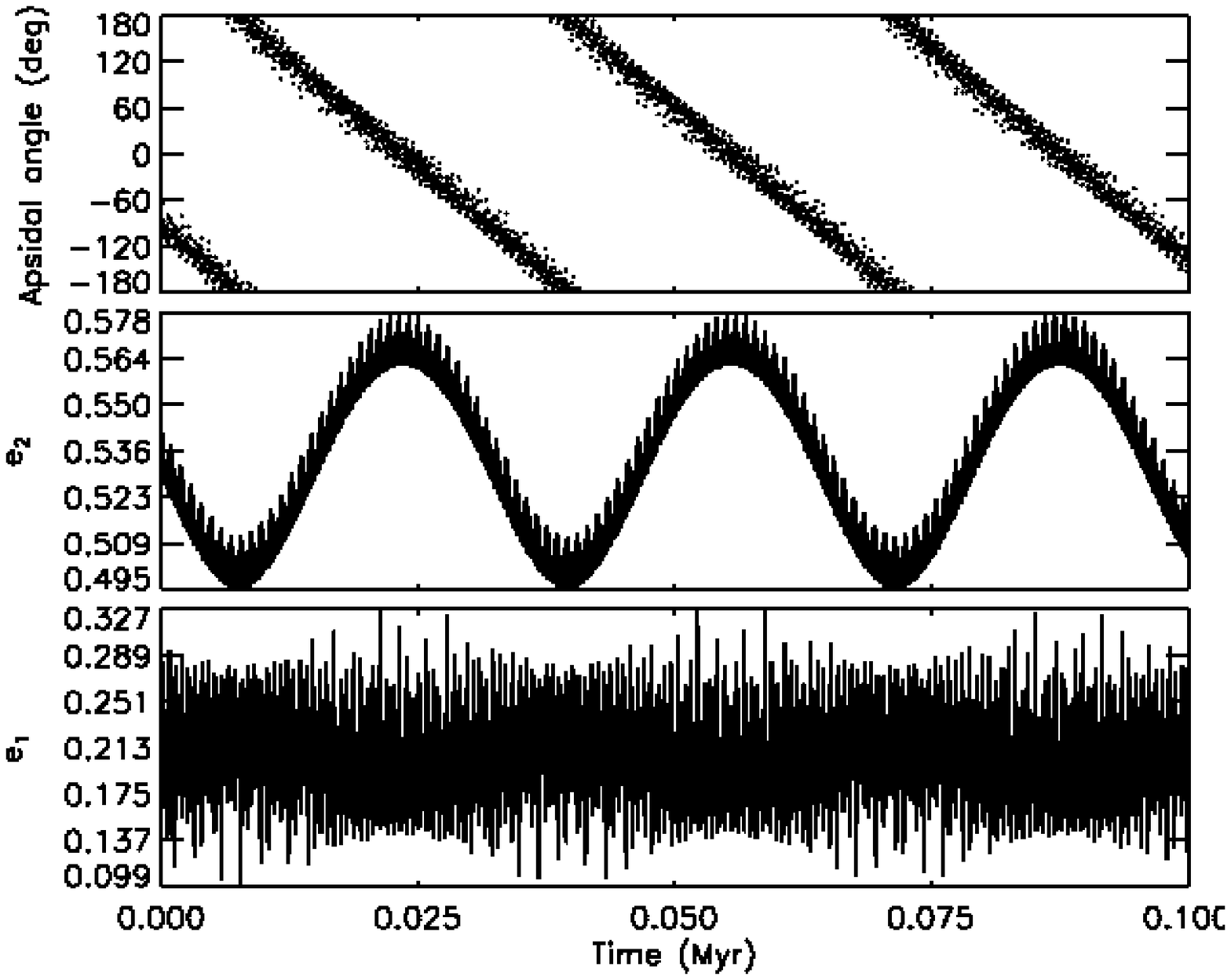,width=\columnwidth,height=3.5truein}}
\figcaption{
Eccentricity and apsidal angle variation from an N-body simulation 
for the HD 168443 system.
\label{plotsix}}
\end{figure}

Although the HD 168443 system \citep{marcyetal01} is thought 
to harbor a brown dwarf \citep{udryetal02,reffert06},
\cite{erdietal04} investigate the dynamical structure of a 
possible habitable zone in the system.  Even though 
HD 168443 b has a large eccentricity ($e_2 \approx 0.53$),
the wide separation of both planets allows the Sundman inequality
to be satisfied.  This same eccentricity, however, produces
a significant difference in the dynamical predictions from
the second- and fourth-order LL theories (Fig. \ref{plotfive}).
Both the fourth-order LL theory and the octupole theory
reproduce the eccentricity values and amplitude of both planets
well, even including the short-period variations in the 
true evolution of $e_1$ (Fig. \ref{plotsix}).
Despite this good agreement, over the course of just $10^5$ yr,
the frequency of the eccentricity variation drifts by about half
of a period.  The circulating apsidal angle observed in the 
true evolution has an additional small-period noisy variation 
which encompasses both the profiles from the fourth-order LL 
theory and the octupole-level theory over $10^5$ yr.

\begin{figure}
\centerline{\psfig{figure=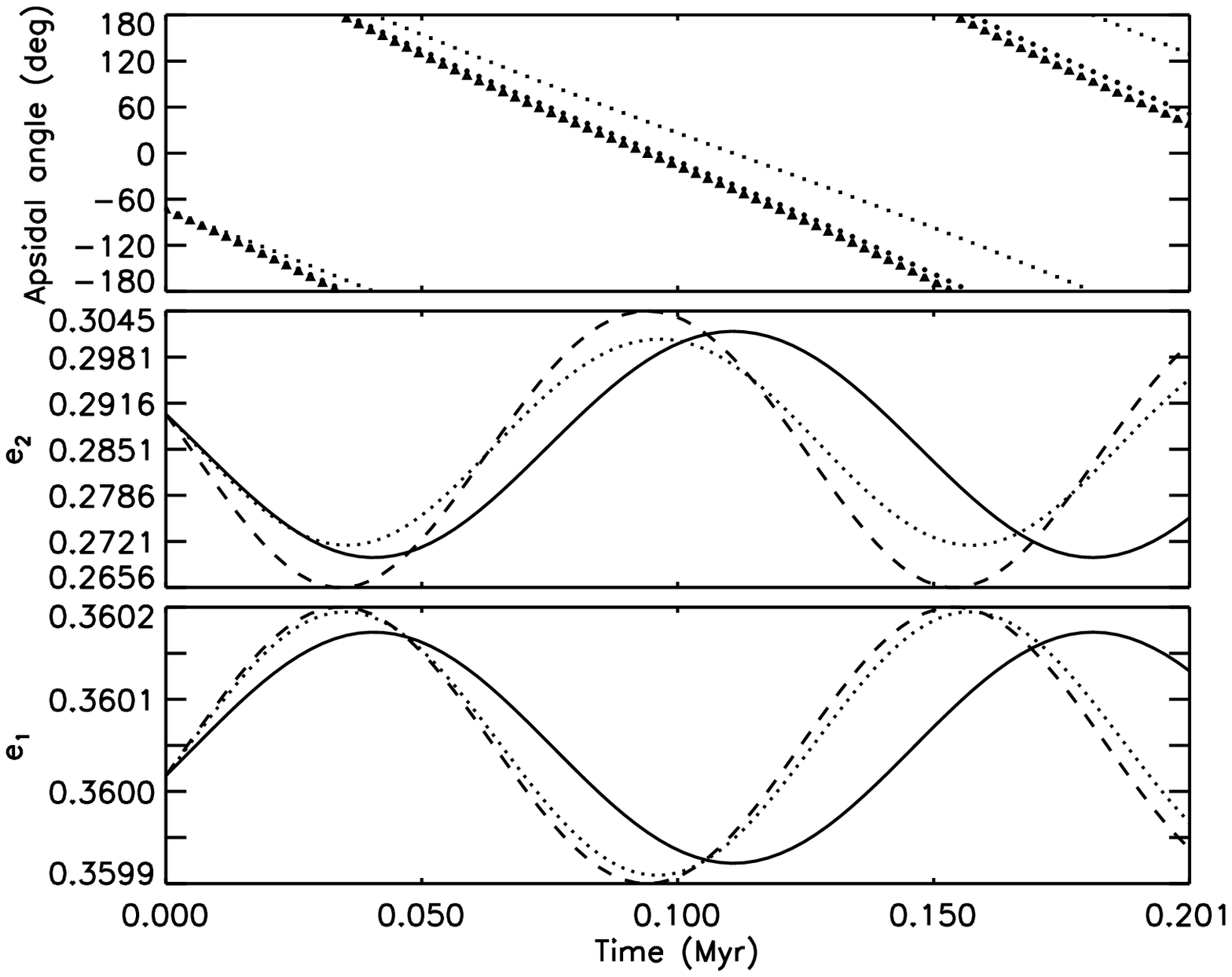,width=\columnwidth,height=3.5truein}}
\figcaption{
Eccentricity and apsidal angle variation according to  
the traditional LL theory (solid lines in the lower two panels and light
squares in the uppermost panel), the fourth-order LL theory 
(dotted lines in the lower two panels and dots in the uppermost panel),
and octupole-level secular perturbation theory 
(dashed lines in lower two panels and thick triangles 
in the uppermost panel) for the HD 38529 system.
\label{plotseven}}
\end{figure}

\begin{figure}
\centerline{\psfig{figure=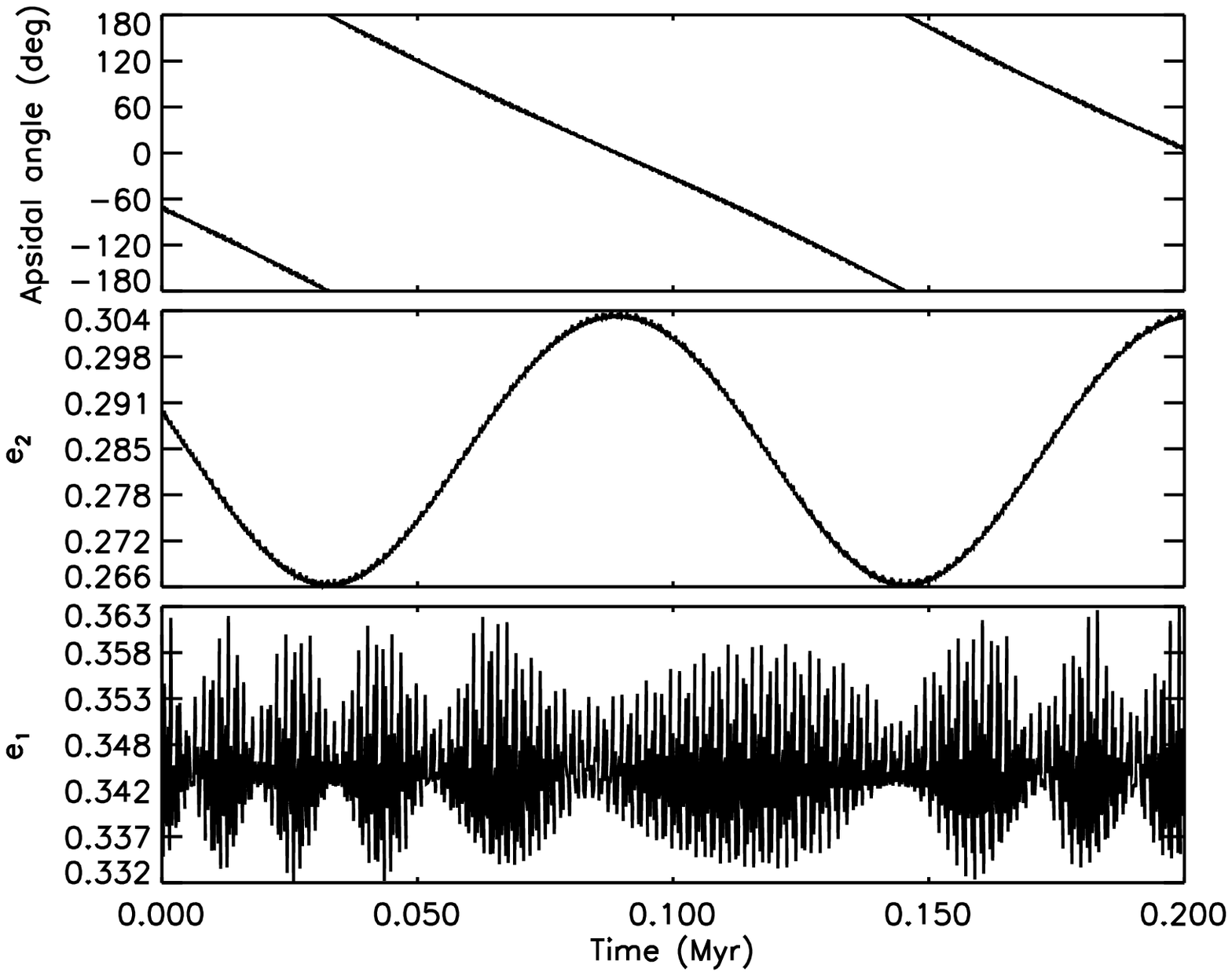,width=\columnwidth,height=3.5truein}}
\figcaption{
Eccentricity and apsidal angle variation from an N-body simulation 
for the HD 38529 system.
\label{ploteight}}
\end{figure}

\begin{figure}
\centerline{\psfig{figure=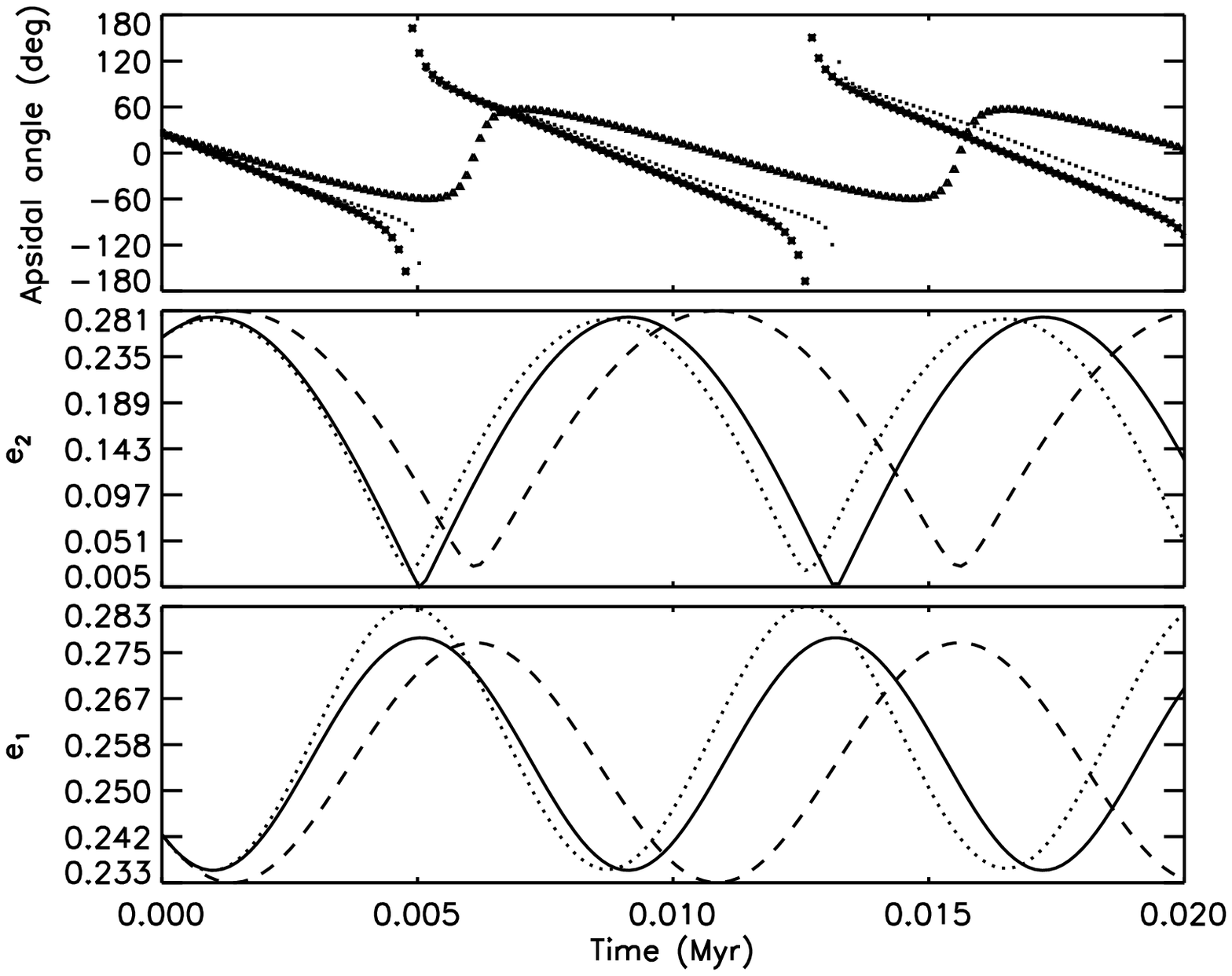,width=\columnwidth,height=3.5truein}}
\figcaption{
Eccentricity and apsidal angle variation according to
the traditional LL theory (solid lines in the lower two panels and light 
squares in the uppermost panel), the fourth-order LL theory 
(dotted lines in the lower two panels and dots in the uppermost panel),
and octupole-level secular perturbation theory 
(dashed lines in the lower two panels and thick triangles
in the uppermost panel) for the planet pair
$\upsilon$ And c and $\upsilon$ And d.
\label{plotnine}}
\end{figure}

\begin{figure}
\centerline{\psfig{figure=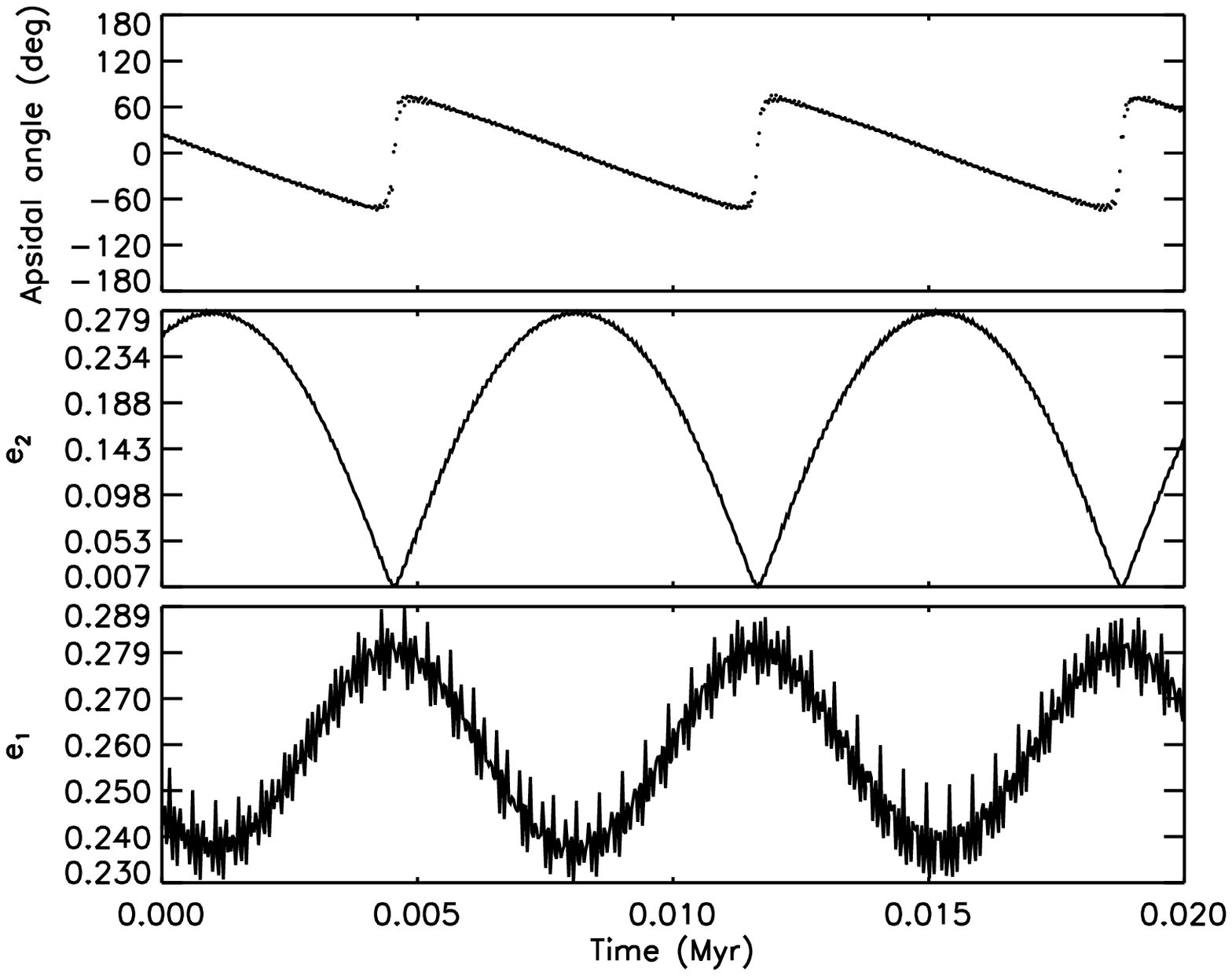,width=\columnwidth,height=3.5truein}}
\figcaption{
Eccentricity and apsidal angle variation from an N-body simulation 
for the planet pair $\upsilon$ And c and $\upsilon$ And d.
\label{plotten}}
\end{figure}

\subsubsection{HD 38529}

The HD 38529 system has been studied in the same context
as the HD 168443 system, as both contain
a planet which could be a brown dwarf.  In HD 38429, however,
the eccentricity of the more massive, outer, planet exceeds that
of the inner planet.  The result of this different configuration
is that octupole theory matches the true evolution better than 
fourth-order LL theory, as can be seen in 
Figs. \ref{plotseven}-\ref{ploteight}.  The
agreement achieved in the eccentricity values and amplitudes
of the inner planet is particularly good (within 1\%) between
the octupole theory and the true evolution.  The eccentricity 
amplitude achieved by the fourth-order LL theory is actually worse
than that from the traditional theory, although the secular
frequency is better modeled by the fourth-order theory.
The long ($> 5 \times 10^4$ yr) circulation period of 
the apsidal angle differs from that predicted by the
fourth-order LL theory and octupole theory by only a few
percent.

Despite some positive agreement, no theory varies the outer
planet's eccentricity by more than $0.0003$, and this eccentricity
is centered nearly around $e_2 = 0.3600$.  Both these characteristics
starkly display the inadequacy of any of these theories to 
quantitatively describe the admittedly minor but non-negligible perturbations
by the less massive inner planet on the more massive moderately eccentric
outer one.

\subsection{Systems with more than two planets}

Although one pair of planets in a multi-planet system may
reside in a mean motion resonance, other pairs of planets in the system 
may evolve primarily by secular means.  In three-planet systems
with a MMR between two of the planets, the other planet's evolution
will be affected, albeit indirectly, by non-secular motions.
Strong (1st, 2nd, or 3rd order) MMRs
are likely to significantly influence the $\mu$ Ara, GJ 876,
and $55$ Cnc systems, and planets in the $55$ Cnc, GJ 876,
$\upsilon$ And, and HD 69830 systems are all close enough 
($a < 0.07$ AU) to their parent stars to have been tidally 
influenced, possibly resulting in a reduction of their 
eccentricities.  Further, a pair of planets in the 
HD 37124 system fail to satisfy the Sundman Criterion
for convergence of Laplace coefficients in terms of the disturbing
function.  All these observations cast doubt on the effective
use of LL secular theory, even a generalized N-body version, 
on the majority of the currently known three and four-planet 
extrasolar planetary systems.  One exception is provided by
\cite{jietal07}, who demonstrate that the
orbital evolution of the three planets in the HD 69830
are well-described with LL secular theory due to their
small eccentricities and relatively large orbital separations.  

\begin{figure}
\centerline{\psfig{figure=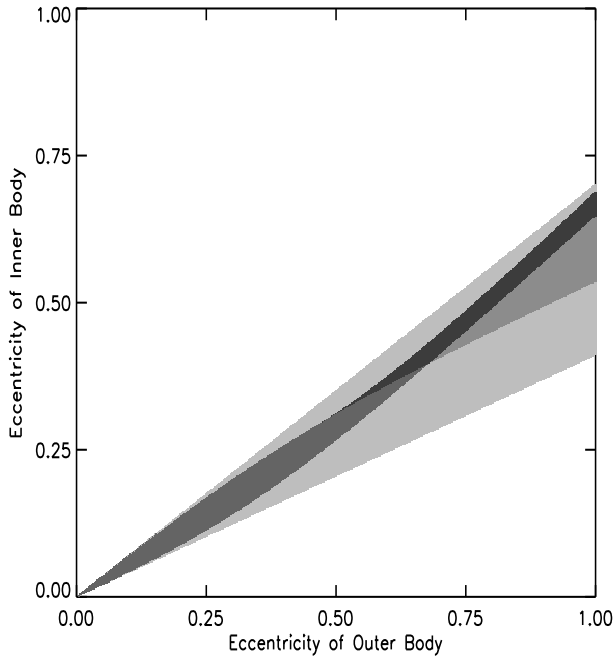,width=\columnwidth,height=3.5truein}}
\figcaption{
Regions of the $\upsilon$ And c - $\upsilon$ And d eccentricity phase space which 
allow for libration from the traditional LL theory only (light gray),
traditional LL theory and the generalized LL theory for the lower sign 
from Eq. (\ref{crit4}) only (medium gray),
traditional LL theory and the generalized LL theory for the upper sign
from Eq. (\ref{crit4}) only (dark gray), 
and the traditional LL theory and both branches of 
the generalized LL theory (black).
\label{ploteleven}}
\end{figure}

Despite these reservations, one may apply LL theory
to the planet pair $\upsilon$ And c and $\upsilon$ And d.
The results, presented in Figs. \ref{plotnine}-\ref{ploteleven},  
perhaps showcase best how even the fourth-order
theory can fail.  Both the traditional LL theory and the
fourth-order theory predict circulation of the apsidal
angle, whereas only the octupole theory instead correctly 
predicts apsidal libration.  This key distinction casts doubt
on the ability of fourth-order LL theory to make reliable
qualitative or quantitative predictions for multi-planet
extrasolar systems in general.

Still, such systems may 
demonstrate the variety of phase space portraits allowed
by the theory.  Figure \ref{ploteleven} displays a panorama
of regions in eccentricity phase space in which apsidal
libration is allowed according to traditional LL theory
and fourth-order LL theory.  The figure hints at the
branch point introduced by the discriminant 
in Eq. (\ref{crit4}), and demonstrates that traditional
LL theory overestimates the region of phase space which
allows for apsidal libration to occur.  This dense and 
varied phase portrait demonstrates
how strongly the fourth-order modification to traditional LL
theory may play a role in an extrasolar system.

\section{Discussion}

Our demonstration of LL theory's limitations for extrasolar
planetary motions of known systems should not overshadow its 
utility in modeling other dynamical systems, including Solar 
System analogs.  Although MMRs help establish the framework of 
planetary systems
\citep{gomesetal05,morbyetal05,tsiganis05}, their final states 
might be dominated
by purely secular perturbations \citep{brouwer50,laskar88,ito02}.  
Quantitatively accurate secular solutions are difficult to obtain, even on 
a $\sim 1-100$ Myr timescale, as both stable and chaotic
long-term ($ \gtrsim 1$ Gyr) solutions exist for the Solar System 
within the uncertainties of current measurements \citep{hayes06}.  
Therefore, analytical results such as those from 
Eqs. (\ref{rate1})-(\ref{rate2}), (\ref{hill})
and (\ref{threephi1})-(\ref{threephi3}) may give 
representative {\it qualitative} descriptions of the 
dynamical state of the system on timescales shorter than those which
may give rise to destructive stochasticity.

A purely secular theory, by definition, does not incorporate
the effects from the resonant terms which
may cause the theory to fail.  The true disturbing function
includes an infinite number of terms, and depending on the
orbital parameters of both resonant bodies, each term
containing mean longitudes has a particular ``strength''.
In essence, a secular theory eliminates all of these terms,
and justifies this neglect by appealing to how the 
combined strength of these terms should not match the strength
of those terms without mean longitudes.  In reality, however,
resonant terms play a non-negligible role in the evolution
of seemingly secular planetary and satellite systems.

\cite{maletal89} presented a way to incorporate ``near-resonant''
terms into Laplace-Lagrange secular theory, significantly
improving their results.  A similar approach could be invoked
for fourth-order LL theory in order to improve its current
general lack of agreement with the actual evolution of multi-planet
exosystems.  If the strong ``near-resonant'' term in question
is a low-order resonance, then only one or two resonant terms
might significantly impact the motion.  With  
assumptions about the period of these terms, their tendency
to circulate instead of librate at near-commensurabilities,
the constancy of their coefficients, and their decoupled states, 
one can model each term as a pendulum, the paradigm for 
single-resonance models. The pendulum model allows one to 
define an energy and compute a circulation period for each
resonant term, and ultimately to include
the relevant averaged effects into the disturbing function.
\cite{maletal89} perform this task for first-order 
near-resonances.  As per their discussion,
expanding their theory in order to encompass a treatment such as
ours would involve deriving the corrections due to 
second-order near-resonances and possibly dealing with the
corrections introduced from coupling between resonant
terms.

Because secular eigenfrequencies are proportional to the
resonant mass/central mass ratio and to the mean motion
of the resonant masses, the utility of LL theory
largely hinges on the values of the masses and semimajor axes
of the resonant objects.
Unlike for secular satellite systems \citep{maletal89,christou97},
the mass ratio in extrasolar systems is orders of magnitude
larger.  The result is that the quality of the results
obtained from LL theory is sensitive to this ratio,
a result also suggested by our numerical investigations.
The planets in HD 108873 are close to a $4$:$1$ MMR.
Using the parameters for the planets from Table 1,
we find that the system's properties are sensitively
dependent on the exoplanet masses and semimajor axis ratio,
but are more robust to changes in initial eccentricity.
A possible (but speculative) reason for this characterization
is that eccentricities appear in the expressions for 
coefficients of resonant terms, but are not included in
the expressions for the secular eignefrequencies.

The neglect of inclination in our fourth-order theory -- although already 
justified by the 1) unknown inclinations of the vast majority of extrasolar 
planets, 2) near-coplanarity of several dynamical subsystems in the 
Solar System, and 3) complexities arising from the coupling of 
eccentricity and inclination
in fourth-order LL theory -- may not be of significant consequence given
the marginal improvement afforded by expanding the theory to
include eccentricity terms up to fourth-order.  However, such coupled terms
might introduce important nonlinearities in the solution of LL theory
for fully three-dimensional problems, and deserve consideration
in future studies.

A subsequent planar expansion 
to arbitrary order is possible and may proceed along the same lines as the
method outlined here.  Expansion to $N$th order will involve terms in the
disturbing function of the form 
$\cos{\left[ N\left( \varpi_l - \varpi_p \right) \right]}$, which can be expressed
as polynomials in $\cos{\varpi_l}$ and $\cos{\varpi_p}$.  The differential
equations representing the motion will contain $h_j$, $h_k$, $k_j$
and $k_k$ terms of all odd orders up to $\left( N-1 \right)$th order.  
However, such an 
expansion may only be useful when seeking a highly accurate 
solution in systems where effects from barycentric drifts, short-period
terms and MMRs may all be safely neglected.

One may also generalize LL theory to fourth-order in the inclinations but
not the eccentricities when modeling inclined circular orbits.  Under
the small angle approximation, $\sin{I_j/2} \approx I_j/2$, where $I_j$
is the inclination of body $j$, the equations
of motions and criteria for nodal libration will have exactly the same form
as those of the planar case, except that the constants will not depend on 
the relative locations of the bodies being perturbed 
 (i.e. no delta functions would appear in the analog of 
Eq. \ref{general}) as there is no preferred reference plane.  A consequence
is that the nodal rate of an inclined massless particle interior to a 
giant planet on a circular coplanar orbit would remain unchanged if the bodies
switched semimajor axes.  To fourth order, this rate would be dependent
on the massless particle's inclination, and be equal to:

\begin{eqnarray}
&&C_{a}^{(j)} \left[
\left( \frac{1}{4} \right) 2 f_{3} + \bigg\lbrace \left( \frac{1}{16} \right) 4 f_8 
- \left( \frac{1}{4} \right) f_3 \bigg\rbrace I_{j}^2
\right]
\nonumber\\
\approx&&
C_{a}^{(j)} \bigg[
-\frac{3}{4} \alpha \left( 1 - \frac{1}{2} I_{j}^2  \right)
+\frac{153}{128} \alpha^2 I_{j}^2
\nonumber\\
&&-\frac{45}{32} \alpha^3 \left( 1 - \frac{1}{2} I_{j}^2 \right)
+\mathcal{O} \left( \alpha^4 \right)
\bigg]
\label{inc}
\end{eqnarray}

Comparison with Eq. (\ref{rate2}) demonstrates a difference
of sign, but a remarkable similarity of coefficients.  In fact,
reduction to the traditional LL theory and to second order in $\alpha$
demonstrates that the nodal rate exceeds the apsidal rate by 
exactly a factor of $\alpha$.  The differences in the presence of 
powers of $\alpha$ in Eqs. (\ref{rate2}) and (\ref{inc}) are due
to the different functions of the Laplace coefficients which arise
naturally from the expansion of Kaula's disturbing function about
zero eccentricities and inclinations.

\section{Conclusion}

We have presented an analytical planar fourth-order (in eccentricity) formulation
of traditional Laplace-Lagrange theory for two non-central bodies 
(Eqs. \ref{eqmo1}-\ref{eqmo} and \ref{twobodya}-\ref{twobodyh}) and 
for $N$ non-central bodies
(Eqs. \ref{eqmo2a}-\ref{eqmo2} and \ref{Nbodya}-\ref{Nbodyk}), and derived conditions for apsidal libration
in various cases (Eqs. \ref{crit2}-\ref{det} and \ref{threephi1}-\ref{RQS}).  We have
surveyed the prospects for secular resonance in multi-planet extrasolar systems, 
and, where appropriate, applied the traditional and generalized
versions of LL theory and octupole-level theory to the systems.  We 
conclude that expansion to high-eccentricity orders fails to 
compensate for the inherent drawbacks of the 
traditional theory when applied to extrasolar planetary systems, and is best
utilized in dynamical systems with more restrictive orbital parameters.

\acknowledgments

We thank Matija \'{C}uk for his useful and spot-on feedback, Renu Malhotra 
for reading the manuscript and for her input, and Eric Ford for a useful 
conversation regarding the idea behind this manuscript.  This work 
was supported by NASA under grants NAG5-13207, NNG04GL01G
and NNG05GI92G from the Origins of Solar Systems, Astrophysics Theory,
and Beyond Einstein Foundation Science Programs respectively, and
by the NSF under grant AST~0407040.

\begin{figure}
\centerline{\psfig{figure=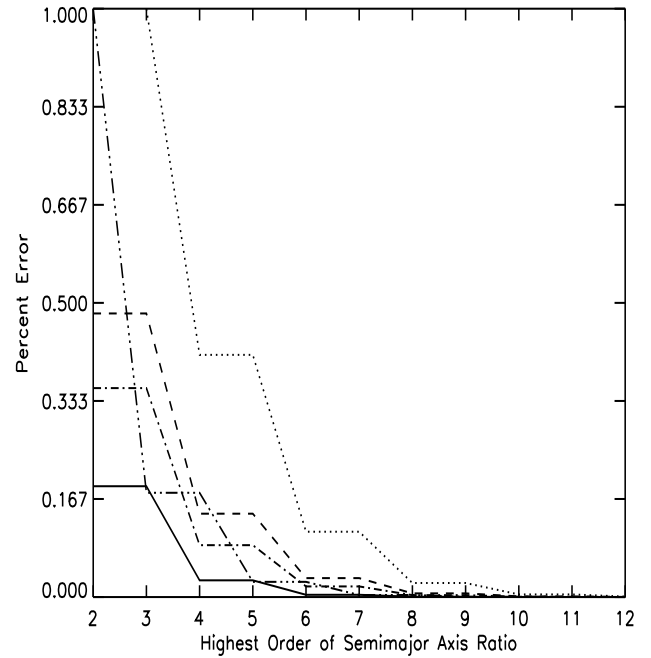,width=\columnwidth,height=3.5truein}}
\figcaption{
Percent error incurred when using the series expansion in $\alpha$
of $f_i$ quantities (in Eqs. B13-B22) relative to the exact 
computation for $f_2$ (solid line), $f_4$ (dotted line), 
$f_5$ (dashed line), $f_6$ (dash-dot line), and 
$f_{10}$ (dash-dot-dot-dot line) for the $\alpha$ given from
HD 12661.
\label{plotthirteen}}
\end{figure}

\appendix
\section{Appendix A}

The following constants in time, expressed as functions
of the planetary semimajor axes and masses only, are used
in the two-body fourth-order LL theory from Eq. (\ref{eqmo}):

\begin{mathletters}
\begin{eqnarray}
&C_{a}^{(1)} \equiv \mathcal{G} a_{1}^{-\frac{3}{2}} m_2 \mu_{1}^{-\frac{1}{2}} \quad\quad
&C_{a}^{(2)} \equiv \mathcal{G} a_{1}^{-1} a_{2}^{-\frac{1}{2}} 
                     m_1 \mu_{2}^{-\frac{1}{2}}
\label{twobodya}
\\
&C_1 \equiv 2 f_2 \quad\quad  
&C_2 \equiv f_{10}
\label{twobodyb}
\\
&C_{3}^{(1)} \equiv 4 f_6 - f_2 \quad\quad 
&C_{3}^{(2)} \equiv 4 f_4 - f_2
\label{twobodyc}
\\
&C_4 \equiv 2 f_5 - 2 f_{17} \quad\quad
&C_5 \equiv 2 f_5 + 2 f_{17}
\label{twobodyd}
\\
&C_{6}^{(1)} \equiv 3 f_{12} - \frac{1}{2} f_{10}  \quad\quad
&C_{6}^{(2)} \equiv 3 f_{11} - \frac{1}{2} f_{10}
\label{twobodye}
\\
&C_{7}^{(1)} \equiv f_{12}  \quad\quad
&C_{7}^{(2)} \equiv f_{11}
\label{twobodyf}
\\
&C_{8}^{(1)} \equiv 2 f_{12}  \quad\quad
&C_{8}^{(2)} \equiv 2 f_{11}
\label{twobodyg}
\\
&C_9 \equiv 4 f_{17} &
\label{twobodyh}
\end{eqnarray}
\end{mathletters}

The following constants are a modified, more general, form of
those in Eqs. (\ref{twobodya})-(\ref{twobodyh}), used in 
the N-body fourth-order 
LL theory from Eq. (\ref{eqmo2}):

\begin{mathletters}
\begin{eqnarray}
C_{a}^{(lp)} &\equiv& 
    \mathcal{G} a_{j}^{-\frac{1}{2}} a_{l}^{-\frac{1}{2}}
     m_k \mu_{j}^{-\frac{1}{2}}
\label{Nbodya}
\\
C_{1}^{(lp)} &\equiv& 2 f_{2}^{(lp)} 
\label{Nbodyb}
\\
C_{2}^{(lp)} &\equiv& f_{10}^{(lp)}
\label{Nbodyc}
\\
C_{3}^{(lp)} &\equiv&  4 \left[ 
           \delta \left( j-p \right) f_{4}^{(lp)} + 
           \delta \left( j-l \right) f_{6}^{(lp)} 
             \right] - f_{2}^{(lp)}
\label{Nbodyd}
\\
C_{4}^{(lp)} &\equiv& 2 f_{5}^{(lp)} - 2 f_{17}^{(lp)}
\label{Nbodye}
\\
C_{5}^{(lp)} &\equiv& 2 f_{5}^{(lp)} + 2 f_{17}^{(lp)}
\label{Nbodyf}
\\
C_{6}^{(lp)} &\equiv&  3 \left[ 
           \delta \left( j-p \right) f_{11}^{(lp)} + 
           \delta \left( j-l \right) f_{12}^{(lp)} 
             \right] - \frac{1}{2} f_{10}^{(lp)}
\label{Nbodyg}
\\
C_{7}^{(lp)}  &\equiv& \delta \left( j-p \right) f_{11}^{(lp)} + 
           \delta \left( j-l \right) f_{12}^{(lp)}
\label{Nbodyh}
\\
C_{8}^{(lp)} &\equiv&  2 \left[ 
           \delta \left( j-p \right) f_{11}^{(lp)} + 
           \delta \left( j-l \right) f_{12}^{(lp)} 
             \right]
\label{Nbodyj}
\\
C_{9}^{(lp)} &\equiv& 4 f_{17}^{(lp)}
\label{Nbodyk}
\end{eqnarray}
\end{mathletters}

\section{Appendix B}

One may express constants found in the coefficients of terms in the
disturbing function as functions of derivatives of Laplace coefficients.
From \cite{murray99}, secular dynamics yield the following forms:

\begin{mathletters}
\begin{eqnarray}
f_2 &=& \frac{1}{4} \alpha \mathcal{D} b_{\frac{1}{2}}^{(0)} \left( \alpha \right)
     + \frac{1}{8} \alpha^2 \mathcal{D}^2 b_{\frac{1}{2}}^{(0)} \left( \alpha \right)
\label{f2a}
\\
f_3 &=& -\frac{1}{4} \alpha b_{\frac{3}{2}}^{(-1)} \left( \alpha \right)
       -\frac{1}{4} \alpha b_{\frac{3}{2}}^{(1)} \left( \alpha \right)
\label{f2b}
\\
f_4 &=& \frac{1}{32} \alpha^3 \mathcal{D}^3 b_{\frac{1}{2}}^{(0)} \left( \alpha \right)
     + \frac{1}{128} \alpha^4 \mathcal{D}^4 b_{\frac{1}{2}}^{(0)} \left( \alpha \right)
\label{f2c}
\\
f_5 &=& \frac{1}{8} \alpha \mathcal{D} b_{\frac{1}{2}}^{(0)} \left( \alpha \right)
     + \frac{7}{16} \alpha^2 \mathcal{D}^2 b_{\frac{1}{2}}^{(0)} \left( \alpha \right)
     + \frac{1}{4} \alpha^3 \mathcal{D}^3 b_{\frac{1}{2}}^{(0)} \left( \alpha \right)
     + \frac{1}{32} \alpha^4 \mathcal{D}^4 b_{\frac{1}{2}}^{(0)} \left( \alpha \right)
\label{f2d}
\\
f_6 &=& \frac{3}{16} \alpha \mathcal{D} b_{\frac{1}{2}}^{(0)} \left( \alpha \right)
     + \frac{9}{32} \alpha^2 \mathcal{D}^2 b_{\frac{1}{2}}^{(0)} \left( \alpha \right)
     + \frac{3}{32} \alpha^3 \mathcal{D}^3 b_{\frac{1}{2}}^{(0)} \left( \alpha \right)
     + \frac{1}{128} \alpha^4 \mathcal{D}^4 b_{\frac{1}{2}}^{(0)} \left( \alpha \right)
\label{f2e}
\\
f_8 &=& \frac{3}{16} \alpha^2 b_{\frac{5}{2}}^{(-2)} \left( \alpha \right)
     + \frac{3}{4} \alpha^2 b_{\frac{5}{2}}^{(0)} \left( \alpha \right)
     + \frac{3}{16} \alpha^2 b_{\frac{5}{2}}^{(2)} \left( \alpha \right)
\label{f2f}
\\
f_{10} &=& \frac{1}{2} b_{\frac{1}{2}}^{(1)} \left( \alpha \right)
        - \frac{1}{2} \alpha \mathcal{D} b_{\frac{1}{2}}^{(1)} \left( \alpha \right)
        - \frac{1}{4} \alpha^2 \mathcal{D}^2 b_{\frac{1}{2}}^{(1)} \left( \alpha \right)
\label{f2g}
\\
f_{11} &=& -\frac{1}{8} \alpha^2 \mathcal{D}^2  b_{\frac{1}{2}}^{(1)} \left( \alpha \right)
        - \frac{3}{16} \alpha^3 \mathcal{D}^3 b_{\frac{1}{2}}^{(1)} \left( \alpha \right)
        - \frac{1}{32} \alpha^4 \mathcal{D}^4 b_{\frac{1}{2}}^{(1)} \left( \alpha \right)
\label{f2h}
\\
f_{12} &=&  \frac{1}{8} b_{\frac{1}{2}}^{(1)} \left( \alpha \right)
     - \frac{1}{8} \alpha \mathcal{D} b_{\frac{1}{2}}^{(1)} \left( \alpha \right)
     - \frac{11}{16} \alpha^2 \mathcal{D}^2 b_{\frac{1}{2}}^{(1)} \left( \alpha \right)
\\
&
     -& \frac{5}{16} \alpha^3 \mathcal{D}^3 b_{\frac{1}{2}}^{(1)} \left( \alpha \right)
     - \frac{1}{32} \alpha^4 \mathcal{D}^4 b_{\frac{1}{2}}^{(1)} \left( \alpha \right)
\label{f2i}
\\
f_{17} &=&  \frac{3}{16} b_{\frac{1}{2}}^{(2)} \left( \alpha \right)
     - \frac{3}{16} \alpha \mathcal{D} b_{\frac{1}{2}}^{(2)} \left( \alpha \right)
     + \frac{3}{32} \alpha^2 \mathcal{D}^2 b_{\frac{1}{2}}^{(2)} \left( \alpha \right)
\\
&
     +& \frac{1}{8} \alpha^3 \mathcal{D}^3 b_{\frac{1}{2}}^{(2)} \left( \alpha \right)
     + \frac{1}{64} \alpha^4 \mathcal{D}^4 b_{\frac{1}{2}}^{(2)} \left( \alpha \right)
\label{f2j}
\end{eqnarray}
\end{mathletters}

\noindent{where} $b_{\frac{1}{2}}^{(0)}$ is a Laplace coefficient
and $\mathcal{D}$ is a differential operator.
Using the hypergeometric expansion of the Laplace coefficients 
\citep{murray99} 
one may show that:

\begin{mathletters}
\begin{eqnarray}
f_2 &=& \frac{3}{8} \alpha^2
     + \frac{45}{64} \alpha^4
     + \frac{525}{512} \alpha^6
     + \frac{11025}{8192} \alpha^8
     + \frac{218295}{131072} \alpha^{10}
     + \frac{2081079}{1048576} \alpha^{12}
\label{lotsoffa}
\\
f_3 &=& -\frac{3}{2} \alpha^2
     - \frac{45}{16} \alpha^4
     -\frac{525}{128} \alpha^6
     - \frac{11025}{2048} \alpha^8
     - \frac{218295}{32768} \alpha^{10}
     - \frac{2081079}{262144} \alpha^{12}
\label{lotsoffb}
\\
f_4 &=& \frac{135}{512} \alpha^4
     + \frac{2625}{2048} \alpha^6
     + \frac{231525}{65536} \alpha^8
     + \frac{1964655}{262144} \alpha^{10}
     + \frac{114459345}{8388608} \alpha^{12}
\label{lotsoffc}
\\
f_5 &=& \frac{9}{16} \alpha^2
     + \frac{225}{64} \alpha^4
     + \frac{11025}{1024} \alpha^6
     + \frac{99225}{4096} \alpha^8
     + \frac{12006225}{262144} \alpha^{10}
     + \frac{81162081}{1048576} \alpha^{12}
\label{lotsoffd}
\\
f_6 &=& \frac{15}{32} \alpha^2
     + \frac{945}{512} \alpha^4
     + \frac{4725}{1024} \alpha^6
     + \frac{606375}{65536} \alpha^8
     + \frac{8513505}{524288} \alpha^{10}
     + \frac{218513295}{8388608} \alpha^{12}
\label{lotsoffe}
\\
f_8 &=& \frac{3}{2} \alpha^2
     + \frac{405}{32} \alpha^4
     + \frac{2625}{64} \alpha^6
     + \frac{385875}{4096} \alpha^8
     + \frac{5893965}{32768} \alpha^{10}
     + \frac{160243083}{524288} \alpha^{12}
\label{lotsofff}
\\
f_{10} &=& -\frac{15}{16} \alpha^3
     - \frac{105}{64} \alpha^5
     - \frac{4725}{2048} \alpha^7
     - \frac{24255}{8192} \alpha^9
     - \frac{945945}{262144} \alpha^{11}
\label{lotsoffg}
\\
f_{11} &=& -\frac{45}{64} \alpha^3
     - \frac{525}{128} \alpha^5
     - \frac{99225}{8129} \alpha^7
     - \frac{218295}{8192} \alpha^9
     - \frac{52026975}{1048576} \alpha^{11}
\label{lotsoffh}
\\
f_{12} &=& -\frac{75}{32} \alpha^3
     - \frac{2205}{256} \alpha^5
     - \frac{42525}{2048} \alpha^7
     - \frac{1334025}{32768} \alpha^9
     - \frac{36891855}{524288} \alpha^{11}
\label{lotsoffi}
\\
f_{17} &=& \frac{315}{256} \alpha^4
     + \frac{4725}{1024} \alpha^6
     + \frac{363825}{32768} \alpha^8
     + \frac{2837835}{131072} \alpha^{10}
     + \frac{156080925}{4194304} \alpha^{12}
\label{lotsoffg}
\end{eqnarray}
\end{mathletters}

Importantly, in the computation of derivatives of Laplacian 
coefficients with negative $j$ values, the absolute value 
of $j$ must be taken {\it before} the application of 
recurrence derivative formulas such as those from 
Eqs. (6.70)-(6.71) of \cite{murray99}.  Fig. 
\ref{plotthirteen} demonstrates the error incurred by 
truncating some of the above expansions to various 
powers of $\alpha$ for $\alpha \approx 0.32$, the value
for the planets in the HD 12661 system.

\section{Appendix C}

Here we express the auxiliary variables from Eq. (\ref{RQS}) 
as time-dependent functions of eccentricities, semimajor axis ratios,
and terrestrial planet-giant planet mass ratio for a giant
planet residing in between (case 1), exterior to (case 2), or
interior to (case 3) two terrestrial planets.  For case 1:

\begin{eqnarray}
Q^{(13)} &=& \mathcal{M} C_{2}^{(13)} \alpha_{12} 
      \left( \sqrt{\alpha_{13}} - 1  \right) e_1 e_3 
     - 2 C_{2}^{(23)} e_2 e_3 \cos{\left( \varpi_1 - \varpi_2 \right)}
\\
S^{(13)} &=& 2 C_{2}^{(23)} e_2 e_3 \sin{\left( \varpi_1 - \varpi_2 \right)}
\\
R^{(13)} &=& 2 \left( \alpha_{12} \sqrt{\alpha_{13}} 
       \frac{C_{1}^{(12)}}{2} e_{1}^2 
      - \frac{C_{1}^{(23)}}{2} e_{3}^2 
      + \alpha_{12} \sqrt{\alpha_{13}} 
        C_{2}^{(12)} e_1 e_2 \cos{\left( \varpi_1 - \varpi_2 \right)} \right)
\end{eqnarray}

\noindent{}For Case 2:

\begin{eqnarray}
Q^{(23)} &=& \mathcal{M} C_{2}^{(23)} 
      \frac{\left( \sqrt{\alpha_{23}} - 1  \right)}
        {\alpha_{12}} e_2 e_3 
     - 2 C_{2}^{(13)} e_1 e_3 \cos{\left( \varpi_1 - \varpi_2 \right)}
\\
S^{(23)} &=& -2 C_{2}^{(13)} e_1 e_3 \sin{\left( \varpi_1 - \varpi_2 \right)}
\\
R^{(23)} &=& 2 \left( \frac{\sqrt{\alpha_{23}}}{\alpha_{12}}
       \frac{C_{1}^{(12)}}{2} e_{2}^2 
        - \frac{C_{1}^{(13)}}{2} e_{3}^2
        + \frac{\sqrt{\alpha_{23}}}{\alpha_{12}}
        C_{2}^{(12)} e_1 e_2 \cos{\left( \varpi_1 - \varpi_2 \right)} \right)
\end{eqnarray}

\noindent{}For Case 3:

\begin{eqnarray}
Q^{(12)} &=& \mathcal{M} C_{2}^{(12)} \alpha_{12} 
      \left( \sqrt{\alpha_{12}} - 1  \right) e_1 e_2 
     - 2 C_{2}^{(23)} e_2 e_3 \cos{\left( \varpi_1 - \varpi_3 \right)}
\\
S^{(12)} &=& 2 C_{2}^{(23)} e_2 e_3 \sin{\left( \varpi_1 - \varpi_3 \right)}
\\
R^{(12)} &=& 2 \left( \alpha_{12} \sqrt{\alpha_{12}} 
           \frac{C_{1}^{(13)}}{2} e_{1}^2 
          - \frac{C_{1}^{(23)}}{2} e_{2}^2 
          + \alpha_{12} \sqrt{\alpha_{12}} 
          C_{2}^{(13)} e_1 e_3 \cos{\left( \varpi_1 - \varpi_3 \right)} \right)
\end{eqnarray}

\end{document}